\documentclass[aps,prl,reprint,superscriptaddress,floatfix,longbibliography]{revtex4-2}

\def\be{\begin{equation}}
\def\ee{\end{equation}}
\def\beq{\begin{eqnarray}}
\def\eeq{\end{eqnarray}}
\def\ba#1{\begin{array}{#1}}
\def\ea{\end{array}}
\def\bn{\begin{enumerate}}
\def\en{\end{enumerate}}

\usepackage{amsmath}
\usepackage{comment}
\usepackage{amsthm}
\usepackage{bbold}
\usepackage{graphicx}
\usepackage{float}
\usepackage{braket}
\newcommand{\oprod}[2]{| #1 \rangle\!\langle #2 |}
\newcommand{\iprod}[2]{\langle #1 | #2 \rangle}

\newcommand{\id}{\mathbb{1}}
\newcommand{\Eps}{\mathcal{E}}
\newcommand{\HS}{\mathcal{H}}
\usepackage[colorlinks=true, allcolors=blue]{hyperref}
\usepackage[titletoc]{appendix}
\usepackage{algorithm}
\usepackage{algpseudocode}

\DeclareMathOperator{\Tr}{Tr}

\usepackage{localdiagrams}
\usepackage{tikz-cd}

\begin{document}

\title{Unveiling Order from Chaos by approximate 2-localization of random matrices}
\author{Nicolas Loizeau}
\affiliation{Department of Physics, New York University, New York, NY, USA}
\author{Flaviano Morone}
\affiliation{Department of Physics, New York University, New York, NY, USA}
\author{Dries Sels}
\affiliation{Department of Physics, New York University, New York, NY, USA}
\affiliation{Center for Computational Quantum Physics, Flatiron Institute, New York, NY, USA}

\date{\today}

\begin{abstract}
Quantum many-body systems are typically endowed with a tensor 
product structure. This structure is 
inherited from probability theory, where the probability 
of two independent events is the product of the probabilities. 
The tensor product structure of a Hamiltonian thus gives a natural decomposition of the system into independent smaller subsystems. 
%Examples of such tensor structures are 1-local (no interactions), 2-local (at most 2 particles interactions), 1D chains, bipartition etc.
Considering a particular Hamiltonian and a particular tensor product structure, one can ask: is there a basis in which this Hamiltonian has this desired tensor product structure? In particular, we ask: is there 
a basis in which an arbitrary Hamiltonian has a 2-local form, i.e. 
it contains only pairwise interactions? 
%In general such an \emph{exact} structure does not exist, however we will 
Here we show, using numerical and analytical arguments, that
generic Hamiltonian (e.g. a large random matrix) can be 
\emph{approximately} written as a linear combination of 
two-body interactions terms with high precision; that is  
the Hamiltonian is 2-local in a carefully chosen basis.
%random matrices are \emph{approximately} 2-local in a carefully chosen basis. 
We show that these Hamiltonians are robust to perturbations. 
Taken together, our results suggest a possible mechanism for the emergence of locality from chaos.
\end{abstract}

% that is to say: is there another Hamiltonian with the same spectrum that has this particular tensor structure?

\maketitle

\section{Introduction}
\begin{comment}
Quantum mechanics is arguably one of the most successful 
theories in physics. It describes the behavior of the 
elementary particles and fundamental forces that make up 
our universe. 
\end{comment}
Typically, to obtain a quantum description of the dynamics 
of a system we go through a procedure of canonical quantization, 
or as Dirac described it~\cite{Dirac1930}, we work by 
\emph{classical analogy}. While this procedure has proven 
extremely powerful, it's also profoundly unsatisfying. 
How can it be that, in order to describe the microscopic 
fundamental quantum laws, we first need to know the 
corresponding classical Hamiltonian that governs the behavior 
of the system? Isn't classical mechanics supposed to emerge out of quantum mechanics? More concretely, since most classical Hamiltonians have a rather simple form, it raises the question on how much we have constrained quantum mechanics by this.  

\begin{comment}It raises some questions regarding the 
completeness of our understanding of quantum mechanics 
itself. Here we investigate this problem by stripping 
quantum mechanics of additional structure, such as the 
Dirac correspondence principle, and asking what physics 
can emerge.
\end{comment}
\begin{figure}
\centering
\includegraphics[width=0.49\textwidth]{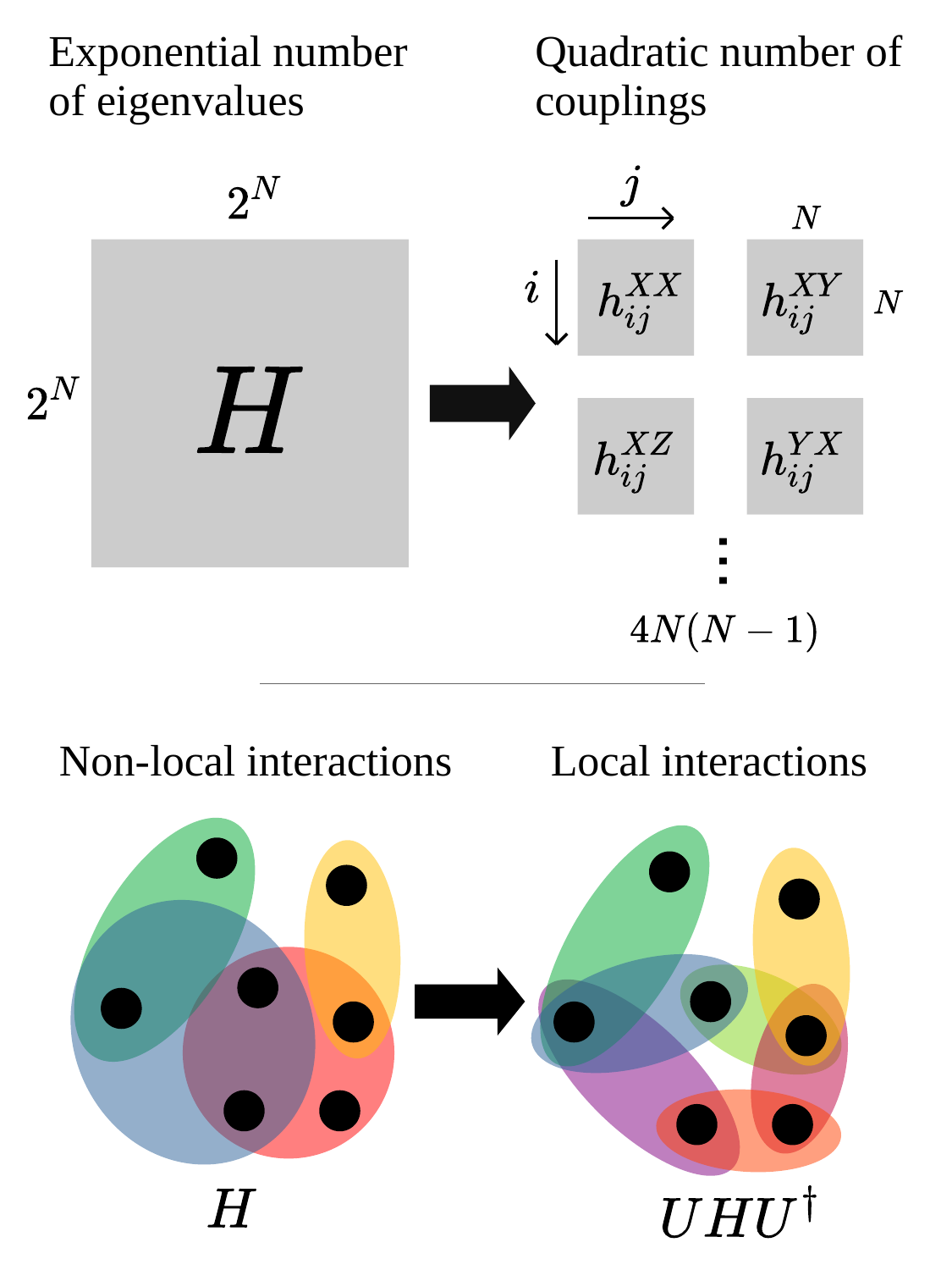}
\caption{\textit{Top}: A $N$ qubit quantum Hamiltonian is a $2^N\times2^N$ hermitian matrix and has $2^N$ intrinsic degrees of freedom (it's spectrum, if a particular basis is specified). After 2-localization, these $2^N$ degree of freedom are compressed into the $O(N^2)$ couplings which specify the 2-local Hamiltonian. \textit{Bottom}: A generic Hamiltonian comprises many body interactions. After 2-localization, the same Hamiltonian is represented in a new basis such that it only consists of 2-body interactions. }
\label{fig:localization}
\end{figure}

To set the stage of our discussion, 
%As a preliminary, 
it's important to elucidate some very basic concepts. We will just reiterate some points made earlier in Ref.~\cite{Sels2014}. 
Quantum mechanics, in and of itself, is independent on one's 
choice of basis, i.e. it is invariant under unitary transformations. In addition, time evolution is a unitary transformation of the state of the system. This puts a constraint on the set of observables 
that can be actually measured. The absence of such 
constraint  would immediately lead to the conclusion that time travel is possible, e.g. instead of measuring observable $O$ one can just measure $\exp(-iHt)O\exp(iHt)$ to travel backwards for time $t$.
Therefore, any discussion should be 
restricted to a specific set of observables. 

In practice, the set of observables we have access to in our 
universe is very limited, and dictated by experimental constraints. Empirically, there is close connection between the Hamiltonian 
and the observables that are accessible; e.g. in quantum field 
theory both typically have simple algebraic expressions in 
terms of creation and annihilation operators~\cite{Weinberg1995}.
Simply put: we write down the Hamiltonian having already 
in mind the observables we're going to measure~\cite{Zanardi2004}. 
It's this tacit assumption of a simple relation between 
the kinematics of the system and the accessible observables 
that we wish to investigate in this work. 

To put it slightly differently, in one of his seminal papers on quantum mechanics Schr\"odinger called entanglement \emph{the} characteristic trait of quantum mechanics that enforces one to depart from classical thinking~\cite{schrodinger_1935}. Entanglement, however, is a basis dependent quantity. It requires one to specify the objects that \emph{naturally} appear as independent, i.e. disentangled. A priori it's not clear what such independent classical objects should be~\cite{Joos1985, Zurek2003, Schlosshauer2005}. Why should one basis be more natural than the other? 

Given a Hamiltonian, the only piece of intrinsic, basis invariant, 
information is its spectrum. Hence, all systems with the same 
energy spectrum are equivalent. The only difference thus hides in 
how we gain access to local physical quantities in those systems. 
This lead to the following question : given a Hamiltonian, is it 
possible to find a basis in which it has a simple (tensor product) 
form, such as a linear combination of only two-body interaction terms? 
The problem has recently been considered by Cotler, Penington and Ranard~\cite{Cotler2019}, who used a simple counting argument to show that it is not possible for most Hamiltonians. However, the question of to what precision it can be done remains open and is the subject of this work. 
We numerically explore a view in which local preferred basis emerges from quantum chaos~\cite{Berry, Haake2001} by looking 
for bases in which random matrices can be approximately written as 2-local Hamiltonians. A related, although quite different, idea has recently been put forward by Freedman and Zini~\cite{Freedman2021a, Freedman2021b} who argue for a novel mechanism of spontaneous symmetry breaking acting on the level of the probability distribution of Hamiltonians rather than on the level of quantum states.

%Entanglement and subsystems depend on the reference frame
% \cite{Ahmad2022,Giacomini2019}. 
%Local hamiltonians \cite{Qi2019}
%Consistent histories \cite{Griffiths1984, GellMann2019}

%Quantum gravity, SYK, black holes and entropy \cite{Bekenstein1973, Maldacena2016, Cotler2017}

\section{2-localization}
\label{sec:2local}
Consider a generic Hamiltonian $H$ that acts on a Hilbert 
space $\HS$. For simplicity let's restrict ourselves to $\HS=\mathbb{C}^M$, where $M$ is taken to be a power of two. To be concrete, think of $H$ as a random  matrix drawn from the GOE ensemble~\cite{Wigner2023, Guhr1998, Atas2013}. In addition, consider the set of Pauli strings $\textsf{P}_N=\{\tau\}$, 
composed out of tensor products of Pauli operators $\sigma_i^\alpha$ 
acting on $N$ spins, or qubits, e.g. 
$\tau=\sigma^x_1 \otimes \sigma^z_2 \otimes \cdots \otimes \mathbb{1}_N$. 
%We define the length of a Pauli string $\tau_i$ to be the number of non-identity operators in the tensor product. 
The set of Pauli strings $\textsf{P}_N$ forms a complete basis, 
hence any Hamiltonian (on $\HS=\mathbb{C}^{2^N}$) 
can be written as a linear combination of Pauli 
strings $H=\sum_{\tau \in \textsf{P}_N} h_{\tau} \tau$. While generic operators are supported on all strings, there is a natural ordering in the set of Pauli strings given by their length, i.e. the number of non-identity operators in the tensor product. Let's denote the set of all strings up to length $k$ as $\textsf{P}^k_N$. In this work we're particularly interested in operators that are localized on the set $\textsf{P}^2_N$ of strings of at most length two. We call a Hamiltonian $2$-localizable if, after some carefully chosen unitary transformation $U$, it is entirely supported on $\textsf{P}^2_N$, i.e. there exists a set of couplings $\{h_\tau\}$ such that:
\begin{equation}
    U H U^\dag = \sum_{\tau \in \textsf{P}^2_N} h_\tau\tau \equiv 
    \sum_{ij, \alpha\beta} J_{ij}^{\alpha\beta} \sigma_i^\alpha \otimes \sigma_j^\beta\ ,
\label{eq:H2local}
\end{equation}
% In particular, a Hamiltonian $H$ is 2-localizable if there exists a unitary $U$ and real coefficients $\{h_{ij}^{\alpha\beta}\}$ such that $UHU^\dag=\sum h_{ij}^{\alpha\beta} \sigma_i^\alpha \otimes \sigma_j^\beta$
where $\sigma_i^\alpha$ is the $\alpha$-Pauli matrix acting on 
the $i$'th qubit.

For example, let's consider a three-qubit problem with a Hamiltonian $H=XXX$, where $X$ stands for $\sigma^x$ and 
the tensor product symbol is dropped for clarity.
This Hamiltonian is 3-local and its spectrum contains an 
equal number of $+1$ and $-1$ eigenvalues. 
But so does the spectrum of $H'=X\id Z$, which is, instead, a 2-local Hamiltonian. Therefore there must exist a unitary $U$ that brings 
$H$ into $H'$ and thus 2-localizes the problem. In fact, it easy to show that $U=\frac{\sqrt 2}{2}(\id\id\id+i\id XY)$ does the job, i.e.
\begin{align}
    U XXX U^\dag &=  X\id Z
\end{align}
%where $X,Y,Z$ are the Pauli matrices and the tensor product symbol 
%is dropped for clarity. 
We can say even more: the Hamiltonian 
$H''=X\id \id$ is 1-local and isospectral to $H$; hence, in this 
simple example, $H$ can be 1-localized. 
In general, the solution is not unique, as one can permute all 
the spins and apply arbitrary single spin rotations. 

A necessary condition for exact 2-localization of an arbitrary matrix $H$ is that there are enough degrees of freedom in the local subspace to encode the eigenvalues of $H$~\cite{Cotler2019}.  There are $3N+\frac{9}{2}N(N-1)$ allowed strings if $H$ is complex (
i.e. drawn from GUE) and $2N+\frac{5}{2}N(N-1)$ if $H$ is real (i.e. drawn from GOE); therefore, the above condition is satisfied for $N\leq8$ in the complex case and $N\leq6$ in the real case, which is consistent with GOE matrices being numerically localizable for $N\leq6$ as we show in the next section. 
Before moving on to the main point of the paper, it's worth noting that this argument has two failure modes: first, it does not say anything about how close one can approximate an operator by a two-local one; second, it does not imply that all operators $N\leq6$ can be 2-localized (it only implies that not all 
%2-local 
Hamiltonians can be 2-localized when $N>6$).
To illustrate the latter, we argue that low rank projectors 
cannot be localized, even in small systems, as we prove next. If a rank-$K$ projector is 2-localizable, then there exists a rank-$K$ projector that is 2-local. So let's derive a bound on the rank of a 2-local projector $P$. $P$ is 2-local iff $P=\sum_{\tau \in 
\textsf{P}^2_N} h_{\tau} \tau$. The rank of $P$ is $K=\Tr(P)=\Tr(P^2)=\sum_{\tau \in \textsf{P}^2_N} \Tr(\tau P)$. Note that $\Tr(\tau P)=2^N h_\tau$, so $K=\frac{1}{2^N} \sum_{\tau \in \textsf{P}^2_N}\Tr(\tau P)^2$. 
Moreover, $P=\sum_q \oprod{q}{q}$, so we have 
\begin{align}
    K&=\frac{1}{2^N} \sum_{\tau \in \textsf{P}^2_N} \left(\sum_q \bra{q}\tau\ket{q}\right)^2\ , \\
    K&\leq \frac{1}{2^N}\mathcal{N}_2K^2 \ ,\\
    \frac{2^N}{\mathcal{N}_2}&\leq K\ ,
\end{align}
where $\mathcal{N}_2=O(N^2)$ is the number of Pauli strings 
of length $2$. 
So any 2-localizable projector on $\mathbb{C}^{2^N}$  has rank greater than $O(2^N/N^2)$. This simply expresses the intuition that one needs non-local information to express a low entropy state $\rho$. 

\section{Method}
\label{sec:method}

There exists a unitary $U$ that localizes a Hamiltonian $H$ if and only if there exists a local Hamiltonian $H'$ that has the same spectrum as $H$. One can localize $H$ by looking for a local Hamiltonian with the same spectrum. 
Let's define the cost function 
\begin{equation}
    C=\frac{1}{2^{N+1}}\sum_{i=1}^{2^N}(E_i-\mathcal{E}_i)^2\ ,
\label{eq:cost}
\end{equation}
where $E_i$ are the eigenvalues of $H$ and 
$\mathcal{E}_i\equiv\mathcal{E}_i(h)$ are the 
eigenvalues of a local Hamiltonian $H'=\sum_{\tau \in \textsf{P}^2_N} h_{\tau} \tau$. 
The cost function $C$ measures the mean squared localization 
error, that is how close the spectrum of the 2-local Hamiltonian 
$H'$ is to the the spectrum of the original Hamiltonian $H$.

Localizing $H$ is equivalent to finding coefficients 
$h_{\tau}$ that minimize $C$. Note that the gradient of $C$ is 
\begin{equation}
\frac{\partial C}{\partial h_\tau}=h_\tau -\frac{1}{2^N}\sum_nE_n \bra{n}\tau \ket{n}
\label{eq:gradient}
\end{equation} 
where $\ket{n}\equiv\ket{n(h)}$ is the eigenvector of $H'$ 
with eigenvalue $\mathcal{E}_n$, as we show explicitly 
in Supplementary Material Sec.3. In practice we minimize $C$ using the Broyden–Fletcher–Goldfarb–Shanno (BFGS) gradient descent method \cite{BROYDEN1970,Fletcher1970,Goldfarb1970,Shanno1970,Fletcher2000}.
In the general case, where the $\tau$ includes $X$,$Y$and $Z$, 
we can 2-localize random matrices $H$ from the GOE ensemble up 
to $N=14$ (i.e. matrices of maximum size $2^{14}\times2^{14}$). 
The main bottleneck is the time required to perform the diagonalization of $H'$ at every step. However, in the case where the $\tau$ are products of $Z$'s only 
(diagonal case) we can 2-localize GOE matrices up to $N=20$. 
When $N>16$ we use a tridiagonal Hermite matrix ensemble to generate the initial GOE spectrum~\cite{Dumitriu2002}.

\section{Results}
\begin{figure}[tb]
\centering
\includegraphics[width=0.49\textwidth]{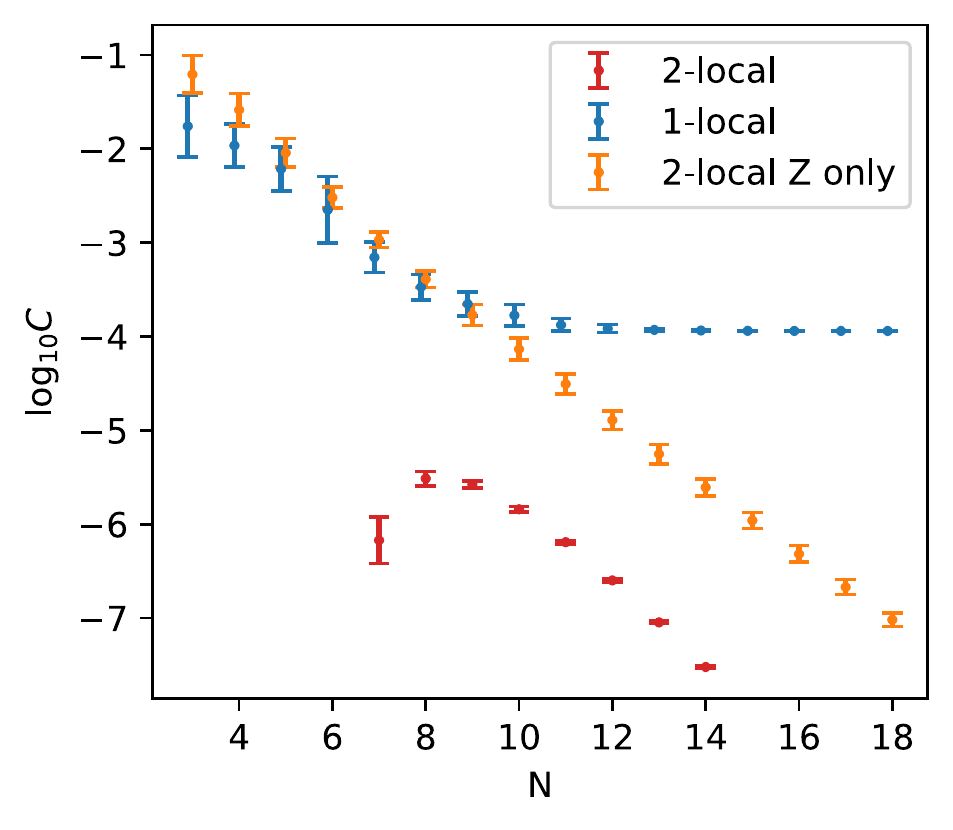}
\caption{Localization error $C$ defined in Eq.~\eqref{eq:cost}
versus system size $N$. Red data points-- labeled ``2 local''-- 
are errors on the localization of a GOE spectrum with 
a 2-local Hamiltonian of the general form given by 
Eq.~\eqref{eq:H2local}. For $N\leq 6$ the error is below machine precision, indicating exact 2-localization, in agreement with the simple counting argument given in the main text. 
For $N>6$ the error vanishes faster than the inverse of the Hilbert space dimension indicating the possibility of 2-localize a 
GOE spectrum with exponential precision.
Blue symbols (``1 local'') are localization errors obtained by using a 1-local Hamiltonian $H'=\sum_ih_i\sigma_i^z$. The latter can be chosen diagonal, since local rotations don't change the $k$-locality. In this case the localization error saturates at $N=10$ and stays constant for 
larger system size, implying the impossibility to 1-localize a 
GOE spectrum. 
Orange points (``2 local  Z only'') corresponds to the case of 
a 2-local Hamiltonian $H'=\sum_{ij}J_{ij}\sigma_i^z\sigma_j^z$, 
describing a classical Ising model. In this case, the localization error decreases 
exponentially with system size, and thus 2-localization is possible in this case. 
Each data point is averaged over 200 GOE matrices. 
%In the 2-local case, the error vanishes like the inverse of the Hilbert space dimension of the system while the error reaches a plateau around $N=10$ in the 1-local case.
}
\label{fig:cost}
\end{figure}

\begin{figure}[tb]
\includegraphics[width=0.49\textwidth]{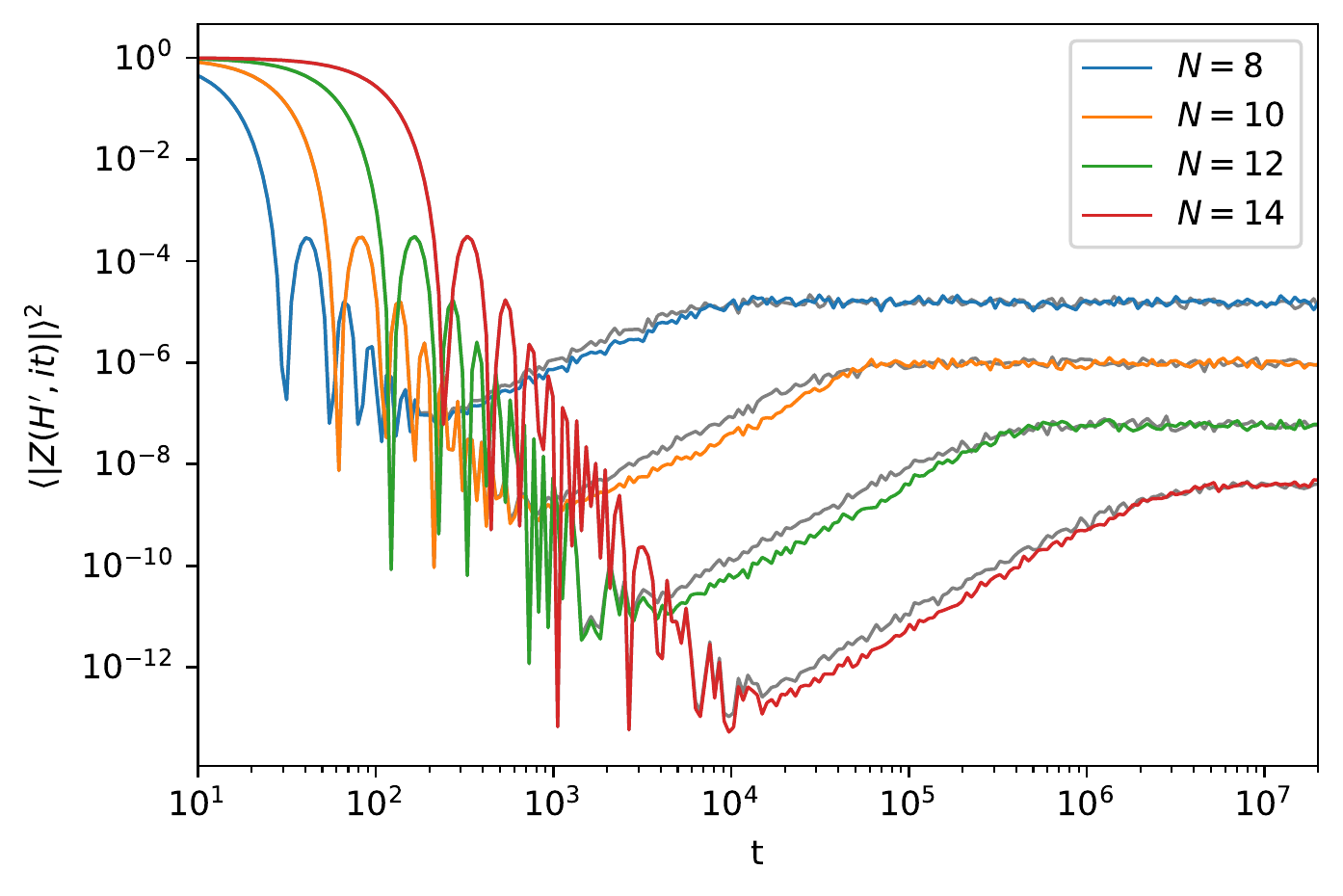}
\caption{Spectral form factor SFF(t) of the localized Hamiltonians (color) defined by Eq.~\eqref{eq:sff}. The SFF of the initial GOE spectrum is plotted in grey for reference. Results are averaged over 200 realizations of GOE initial spectra.}
\label{fig:SFF}
\end{figure}

We generate sets of 200 $N$-qubits random Hamiltonians from the 
Gaussian orthogonal ensemble (GOE); then attempt to 2-localize them. 
For $N\leq6$ every particular Hamiltonian $H$ was localizable up 
to machine precision. For $N>6$, the larger the system, the 
better they can be localized and the error decreases exponentially  
with $N$ as shown in figure \ref{fig:cost}, i.e. in the accessible regime the error goes down faster than the inverse dimension of Hilbert space $2^{-N}$.
Since we retrieve the spectrum with exponential precision it seems likely we do not just retrieve coarse grained information about the density of states of $H$, but reproduce all essential features. To verify this we compare the spectral form factor (SFF) of the retrieved ensemble of 2-local $H'$ with that of the GOE ensemble. The SFF can be thought of as the Fourier transform of the two-point correlation function of the spectrum, i.e. it measures how fluctuations in the density of states are correlated
\begin{equation}
    {\rm SFF}(t)= \left< |Z(H,it)|^2 \right>,
\label{eq:sff}
\end{equation}
where $Z(H,it)$ denotes the generating function  
\begin{equation}
Z(H,it)=\Tr e^{itH},   
\end{equation}
and the $\left< \cdot \right>$ refers to the ensemble average 
over $H$. The spectral form factor has a universal ramp structure at late times which is a hallmark of quantum chaos~\cite{Mehta2004,Haake2001}, see Fig.~\ref{fig:SFF}. In typical chaotic systems, there is some non-universal initial 
behavior ending in the so-called correlation hole, the time duration of which is sometimes called the Thouless time.
%The onset of this universal behavior is sometimes called the Thouless time,after which one observes the so-called correlation hole. 
After the Thouless time follows a universal ramp which stops 
at the Heisenberg time. The study of this universal behavior has yielded important insights into ergodicity breaking, in particular in the context of disordered many-body systems~\cite{lev2020,jed2021} and SYK models~\cite{Cotler2017,Maldacena2016}.
As shown in Fig.~\ref{fig:SFF}, we recover all essential features of the SFF in the 2-localized ensemble at all timescales. We only observe a small
%, both clear, 
deviation in the ramp which can be interpreted as a small delay in the Thouless time.

\subsection{Stability}
Having established that there are 2-local Hamiltonians that approximate a GOE matrix with exponential precision, it becomes important to understand the stability of these solutions. If small changes in the coupling constants $h_\tau$ result in a completely different spectrum this would make the 2-local Hamiltonians rather fine-tuned. Consider the Hessian of the cost function $C(h)$ in the minimum $h=h^0$ (see Supplementary Material Sec.3 for details):
\begin{equation}
    g_{\tau \eta}(h_0) = \frac{\partial^2 C}{\partial h_\tau \partial h_\eta}\Big|_{h=h^0} = 
    \frac{1}{2^N}\sum_{n}\bra{n} \tau \ket{n}\bra{n}\eta \ket{n}.
    \label{eq:Hessiandef}
\end{equation}
where $\ket{n}\equiv\ket{n(h)}$ is the eigenvector of $H'$. Note that the Hessian only depends on the diagonal expectation values of 2-local operators, which are expected to behave completely differently in integrable and chaotic systems~\cite{rigol16review}. In that regard, consider a 2-local $H'$ in which the Pauli strings are restricted to commute, e.g. strings composed of only $Z$'s, which are all diagonal. All eigenvectors of $H'$ are eigenvectors of $\tau$, such that the sum in expression~\eqref{eq:Hessiandef} becomes a trace. Since all strings are trace orthogonal, one finds $g_{\tau \eta}=\delta_{\tau\eta}$. 
As a consequence, for commutative 2-local Hamiltonians, a small change in the coupling constants $\Delta h$ results in a change of the cost function $\Delta C\approx \Vert \Delta h\Vert^2$. Since the coupling constants themselves are $O(1/N)$ this requires exponential precision in the specification of the coupling constants $h$ to maintain the exponential decrease in the cost 
function seen in  Fig.~\ref{fig:cost}. Also, since the metric $g_{\tau \eta}$ becomes diagonal there are no particular directions of stability:  the system is equally susceptible to small perturbations in all directions. 

\begin{figure}[tb]
\includegraphics[width=0.49\textwidth]{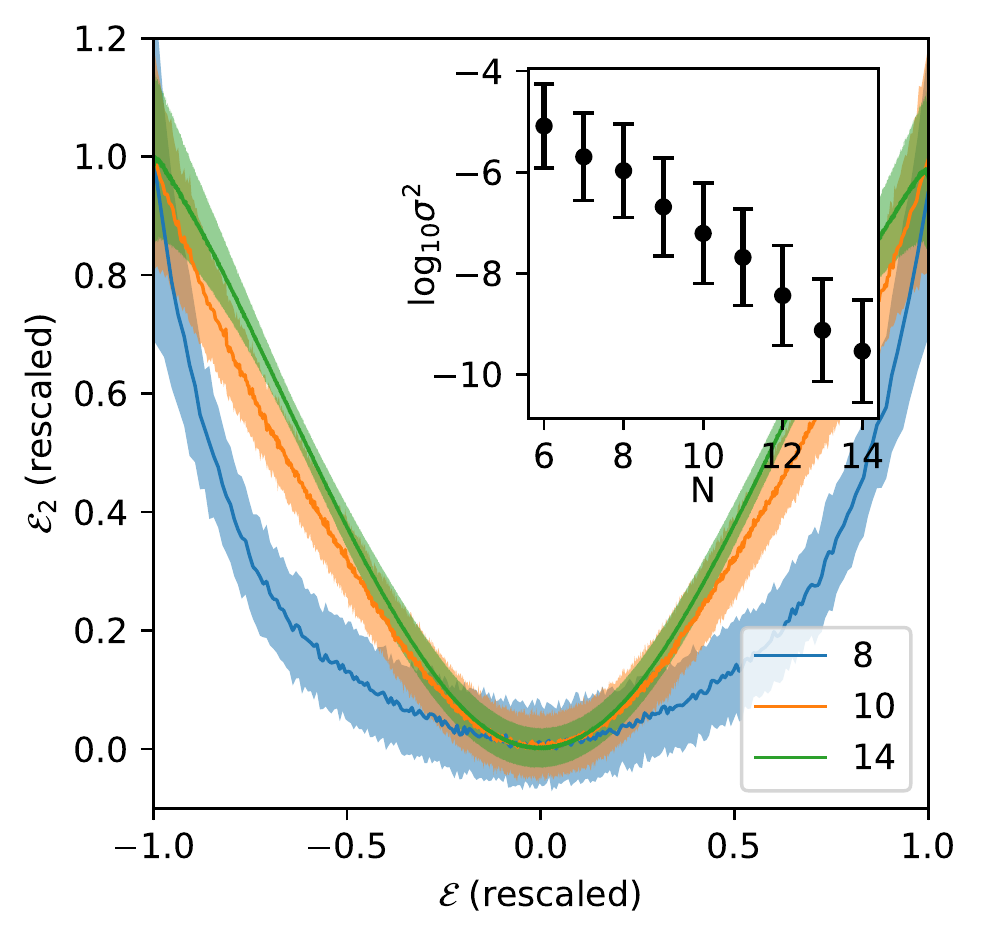}
\caption{Expectation value of the second eigenoperator of $g_{\tau \eta}$ defined by expression~\eqref{eq:Ok} in the eigenstates of the 2-local Hamiltonian $H'$. The figure shows $\left<n|O_2|n\right>$ as a function of the eigenenergy $\mathcal{E}_n$. The line is the mean over the 200 realizations. The shaded region is the standard deviation. The sub-plot shows the residual from a 6th order polynomial fit to the data, which is used to subtract the smooth part of the result. One clearly observes an exponential suppression of the fluctuations with system size.}
\label{fig:smoothness_std}
\end{figure}

The situation should be different for generic $H'$ in which the diagonal expectation values in expression~\eqref{eq:Hessiandef} are expected to obey the eigenstate thermalization hypothesis (ETH)~\cite{rigol16review}. According to ETH, expectation values of (local) observables become smooth functions of energy which drastically alters the behavior of the metric $g_{\tau \eta}$. 
To verify this hypothesis we need first to note that  
the eigenvectors of $g_{\tau \eta}$, denote them by $v^k$, 
are dual to operators $O_k$, defined as:
\begin{equation}
O_k = \sum_{\tau \in \textsf{P}^2_N} v^k_\tau\tau\ . 
\label{eq:Ok}
\end{equation}
Numerical diagonalization of the metric indeed confirms that operators $O_k$ have smooth expectation values in the energy eigenbasis of $H'$. For example, Fig.~\ref{fig:smoothness_std} depicts the behavior of the expectation values $\mathcal{E}_2=\left<n|O_2|n\right>$ of the 
operator $O_2$ corresponding to second eigenvector $v^2$, as a function of energy $\mathcal{E}$, showing that 
$\mathcal{E}_2(\mathcal{E})$ becomes a smooth function of 
$\mathcal{E}$ with increasing system size 
$N$. 

The functional behavior of eigenoperator is also rather simple, which begs the question of whether we can understand the spectrum of $g_{\tau \eta}$ in more details.  
First of all, it is easy to check that $h^0$ is an eigenvector of $g_{\tau \eta}$ 
corresponding to the largest eigenvalue $\lambda_1=1$:
\begin{align}
    \sum_{\eta}g_{\tau \eta} h^0_\eta &= 
    \frac{1}{2^N}\sum_{n}\bra{n}\tau \ket{n}\bra{n} \sum_{\eta \in \textsf{P}^2_N}h^0_\eta \eta \ket{n}  \nonumber \\
    &= \frac{1}{2^N}\sum_{n}\bra{n}\tau \ket{n}\bra{n}H\ket{n}  \nonumber \\
    &= \frac{1}{2^N}\sum_{n}E_n\bra{n}\tau \ket{n} = h^0_\tau \ ,
\label{eq:v1}
\end{align}
where, in the last step, we used the fact the gradient in Eq.~\ref{eq:gradient} is zero when evaluated in $h^0$. This means that a perturbation in the direction of $h^0$ 
increases the cost function significantly. However, the associated operator $O_1$ is just the Hamiltonian $H'$ itself. Such perturbations thus only result in a rescaling of the Hamiltonian. It's obvious why this increases the cost $C$ but since this can just be absorbed in a redefinition of time it leaves all the physics invariant. 
It's more interesting to understand the behavior of the sub-leading eigenvalues.

\begin{figure}[tb]
\includegraphics[width=0.49\textwidth]{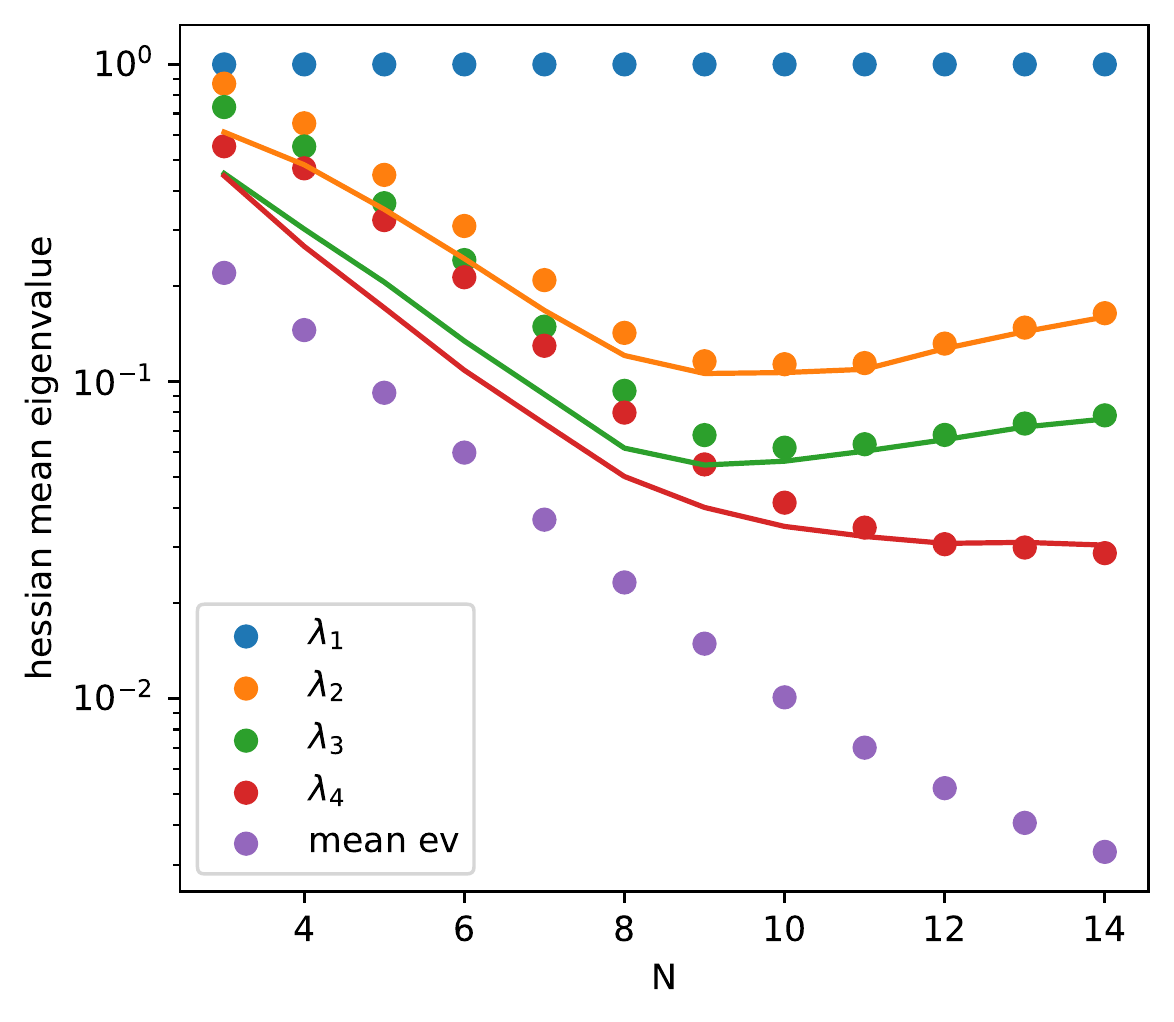}
\caption{The four largest eigenvalues $\lambda_k$ of the hessian $g_{\tau \eta}$ as a function of system size $N$. 
The largest eigenvalue $\lambda_1=1$ corresponds to 
the eigenvector $v^1\equiv h^0_\tau$ (see Eq.~\eqref{eq:v1}). 
The subsequent eigenvalues are all smaller than one. Their 
large $N$ behavior is especially important to understand 
how fine-tuned the coupling constants $J_{ij}^{\alpha \beta}$ 
of the 2-local Hamiltonian $H'$ are(i.e., the ``sloppiness'' 
of the 2-local Universe ); hence their asymptotic behavior 
is discussed extensively in the main text, where we suggest 
a plausible scenario and its implications. 
Each dot correspond to an average over 200 realizations. Solid lines are estimates using expression~\eqref{eq:lambda_k} where $F_k(H)$ is constructed using the procedure explained in the 
main text.
%with the corresponding orthogonal polynomials of degree $n$.  
}
\label{fig:spectrumHess}
\end{figure}

In general, the full set of eigenvalues of $g_{\tau \eta}$ is given by (see Supplemetary Material Sec. 4)
%\ref{sup:hessianev}):
\begin{equation}
    \lambda_k = \frac{1}{2^N}\frac{ \sum_{\tau \in \textsf{P}^2_N} \Tr\left(F_k(H')\tau\right)^2}{\Tr\left(F_k(H')^2\right)}\ ,
    \label{eq:lambda_k}
\end{equation}
where $F_k(H')$ is a function of 2-local $H'$. In general $F_k$ need not be well behaved, like in the diagonal case described earlier. However, assuming the eigenstates of $H'$ obey ETH these functions should be smooth. We've already established that $F_1(x)=x$ and we know all operators need to be traceless. Furthermore, the eigenvalues, given by eq.~\eqref{eq:lambda_k} have a simple interpretation. They are the square Frobenius norm of (normalized) projection of $F_k(H')$ on the two local subspace $\textsf{P}^2_N$. Powers of $H'$ generate more and more non-local strings, which suggest that the eigenoperators are close to projected orthogonal polynomials of $H'$. We indeed find that the $F_k$ can be very well approximated by Gram-Schmidt 
orthogonalization of polynomial function of $H'$ of degree $k$, starting with $F_{0}(x)=\id$ (to ensure tracelessness) and $F_1(x)=x$. Thus, for example, $F_2(x)$ is the traceless part of $x^2$. i.e. 
\begin{equation}
    F_2(x) = x^2 - \frac{\Tr(x^2)}{2^N}\ .
    \label{eq:F2}
\end{equation}
The results for the first few eigenvalues of $g_{\tau \eta}$ are shown in Fig.~\ref{fig:spectrumHess}, together with the exact eigenvalues. The largest eigenvalue is indeed 1 and all the other eigenvalues decay rapidly with $N$ for small systems. Some of the large eigenvalues appear to recover or slow down for larger systems. Note that we find excellent agreement between the approximate eigenvalues constructed from $F_k$ and the exact results at larger $N$. In addition, these eigenvalues are variational so they form a lower bound to the true eigenvalues. 

In principle, closed form expression for the eigenvalues can be extracted exactly from expression~\eqref{eq:lambda_k} in terms of the coupling constants $h$; the problem is entirely algebraic. Nonetheless, the general problem is rather cumbersome to say the least. To make further progress we restrict ourselves to compute $\lambda_2$ from Eqs.~\eqref{eq:F2} and~\eqref{eq:lambda_k} under the assumption that the model is diagonal, i.e. $H'$ is an Ising model with coupling $J$ (we verify in the Supplemetary Material Sec. 5 and 6 and Fig. S2 that this approximation does not affect the qualitative behavior). A detailed diagrammatic calculation is presented in the Supplemetary Material Sec. 5, and the result reads:
%We can use equation\eqref{lambda_k} to show that in the case the $\tau_i$ are all diagonal, the second eigenvalue is
\begin{align}
    &\lambda_2 =\nonumber\\
    &=\frac{ 2\Tr(J^4)-\sum_{i}\left[(J^2)_{ii}\right]^2}
    { 3 \Tr J^4+\frac{1}{2}\Tr(J^2)^2 -6\sum_{i}\left((J^2)_{ii}\right)^2+2\sum_{ij}(J_{ij})^4}\ ,
    \label{lambda2_approx}
\end{align}
%where $\alpha=-\sum_{i}\left[(J^2)_{ii}\right]^2$ , $\beta=-6\sum_{i}\left((J^2)_{ii}\right)^2+2\sum_{ij}(J_{ij})^4$
%and $J$ is the coupling matrix i.e $H=\sum_{ij}J_{ij}Z_iZ_j$.
which, for large $N$, becomes
\begin{align}
    \lambda_2 \sim\frac{2 \Tr(J^4)}{ 3 \Tr J^4+\frac{1}{2}\Tr(J^2)^2}
    \leq 4 \frac{ \Tr(J^4)}{\Tr(J^2)^2}.
    \label{eq:bound2}
\end{align}
The latter can be interpreted as the inverse participation ratio or purity of the spectrum of the matrix $J^2$~\cite{Bell1970}.  
Thus, if all eigenvalues $\mu_{J}$ participate equally to $\lambda_2$ 
then $\lambda_2 = O(N^{-1})$;
%and the corresponding eigenvector is {\it extended}, that is has a finite fraction of non-zero components;
on the contrary, if only few of the $\mu_J$'s participate, then 
$\lambda_2 = O(1)$. This remark is of particular importance in that it 
may explain the emergence of 2-locality altogether. If $\lambda_2 \rightarrow 0$ then \emph{any} perturbation of the coupling constants $h$ is marginally irrelevant, meaning that it won't change the spectrum of the theory when $N \rightarrow \infty$. These coupling constants are thus by no means fined tuned to any specific value: they just happen to have a particular value, 
but the volume of allowed values they can take on, 
leaving the physics unchanged, is enormous. 
Alternatively $\lambda_2$ tends to a constant in the thermodynamic limit, which means that $J$ can be approximated by a low rank matrix. The numerical data in Fig.~\ref{fig:spectrumHess} suggest this might be the case.  
Arguments can be given either way, on the one hand it seems expected that one has to put a few more constraints, other than the bandwidth to stay close to the desired density of states. On the other hand, we've also explicitly computed the cost function for low-rank $J$ in the Supplementary Material Sec. 6 (see also Fig. S3) which suggests the cost function saturates at a constant at fixed rank, implying finite error for localization for large $N$.
Regardless of the final outcome, our results lead to the remarkable observation that one can either 2-localize a GOE random matrix on a model with a finite number of parameters or there is large emergent invariance. 

Before we conclude, let's stress that, even when $\lambda_2$ is nonzero, 2-local Hamiltonians are expected to be very sloppy~\cite{sethna13}. That is, most combinations of parameters $h$ are unimportant. The conclusion follows from a generalization of expression~\eqref{eq:bound2} to higher $k$. While it's rather cumbersome to establish the full result, the leading order contribution behaves like
$\lambda_k \sim \Tr(J^{2k})/\Tr(J^2)^k$ where a careful analysis suggests that the multiplicity of diagrams contributing to the numerator and denominator are the same. As a result the 
eigenvalues of the metric are expected to follow a 
geometric progression, 
%geometrically distributed, 
the hallmark of ``sloppiness''~\cite{sethna13}.  

\section{Conclusion and discussion}
We find that random GOE matrices can be represented in a local form with a very good precision, i.e. the norm of the remaining non-local part decreases exponentially with the size of the system. This effectively corresponds to an exponential compression of the amount of data. Among other things, this is a step toward the resolution of the preferred basis problem: associated with each random Hamiltonian, there is a preferred basis in which this Hamiltonian has an almost local description. While the present work does not require geometric locality, i.e our local Hamiltonians can represent all-to-all particles interactions, the sloppiness of the program suggests the model can still be greatly simplified without fine tuning. This suggest a route to understand how space could emerge from quantum mechanics alone by looking at the adjacency structure of the couplings $J_{ij}$~\cite{Cao2017,Carroll2021,Bekenstein1973, Maldacena2016, Cotler2017} . On the other hand, we also showed that even if generic random Hamiltonians can be localized efficiently, some particular Hamiltonians cannot. These are examples of operators that have some fundamental quantum non-local properties and cannot be represented as a sum of 2-body operators.

\emph{Acknowledgments.} The Flatiron Institute is a division of the Simons Foundation. We acknowledge support from AFOSR: Grant FA9550-21-1-0236.

\bibliography{bib}

\begin{thebibliography}{33}%
\makeatletter
\providecommand \@ifxundefined [1]{%
 \@ifx{#1\undefined}
}%
\providecommand \@ifnum [1]{%
 \ifnum #1\expandafter \@firstoftwo
 \else \expandafter \@secondoftwo
 \fi
}%
\providecommand \@ifx [1]{%
 \ifx #1\expandafter \@firstoftwo
 \else \expandafter \@secondoftwo
 \fi
}%
\providecommand \natexlab [1]{#1}%
\providecommand \enquote  [1]{``#1''}%
\providecommand \bibnamefont  [1]{#1}%
\providecommand \bibfnamefont [1]{#1}%
\providecommand \citenamefont [1]{#1}%
\providecommand \href@noop [0]{\@secondoftwo}%
\providecommand \href [0]{\begingroup \@sanitize@url \@href}%
\providecommand \@href[1]{\@@startlink{#1}\@@href}%
\providecommand \@@href[1]{\endgroup#1\@@endlink}%
\providecommand \@sanitize@url [0]{\catcode `\\12\catcode `\$12\catcode
  `\&12\catcode `\#12\catcode `\^12\catcode `\_12\catcode `\%12\relax}%
\providecommand \@@startlink[1]{}%
\providecommand \@@endlink[0]{}%
\providecommand \url  [0]{\begingroup\@sanitize@url \@url }%
\providecommand \@url [1]{\endgroup\@href {#1}{\urlprefix }}%
\providecommand \urlprefix  [0]{URL }%
\providecommand \Eprint [0]{\href }%
\providecommand \doibase [0]{https://doi.org/}%
\providecommand \selectlanguage [0]{\@gobble}%
\providecommand \bibinfo  [0]{\@secondoftwo}%
\providecommand \bibfield  [0]{\@secondoftwo}%
\providecommand \translation [1]{[#1]}%
\providecommand \BibitemOpen [0]{}%
\providecommand \bibitemStop [0]{}%
\providecommand \bibitemNoStop [0]{.\EOS\space}%
\providecommand \EOS [0]{\spacefactor3000\relax}%
\providecommand \BibitemShut  [1]{\csname bibitem#1\endcsname}%
\let\auto@bib@innerbib\@empty
%</preamble>
\bibitem [{\citenamefont {Dirac}(1930)}]{Dirac1930}%
  \BibitemOpen
  \bibfield  {author} {\bibinfo {author} {\bibfnamefont {P.~A.~M.}\
  \bibnamefont {Dirac}},\ }\href@noop {} {\emph {\bibinfo {title} {The
  principles of quantum mechanics}}},\ \bibinfo {edition} {4th}\ ed.\ (\bibinfo
   {publisher} {Oxford},\ \bibinfo {year} {1930})\BibitemShut {NoStop}%
\bibitem [{\citenamefont {Sels}\ and\ \citenamefont
  {Wouters}(2014)}]{Sels2014}%
  \BibitemOpen
  \bibfield  {author} {\bibinfo {author} {\bibfnamefont {D.}~\bibnamefont
  {Sels}}\ and\ \bibinfo {author} {\bibfnamefont {M.}~\bibnamefont {Wouters}},\
  }\bibfield  {title} {\bibinfo {title} {Quantum equivalence, the second law
  and emergent gravity},\ }\bibfield  {journal} {\bibinfo  {journal} {arxiv}\
  }\href {https://doi.org/10.48550/arXiv.1411.3901} {10.48550/arXiv.1411.3901}
  (\bibinfo {year} {2014})\BibitemShut {NoStop}%
\bibitem [{\citenamefont {Weinberg}(2005)}]{Weinberg1995}%
  \BibitemOpen
  \bibfield  {author} {\bibinfo {author} {\bibfnamefont {S.}~\bibnamefont
  {Weinberg}},\ }\href {https://doi.org/10.1017/CBO9781139644167} {\emph
  {\bibinfo {title} {{The Quantum theory of fields. Vol. 1: Foundations}}}}\
  (\bibinfo  {publisher} {Cambridge University Press},\ \bibinfo {year}
  {2005})\BibitemShut {NoStop}%
\bibitem [{\citenamefont {Zanardi}\ \emph {et~al.}(2004)\citenamefont
  {Zanardi}, \citenamefont {Lidar},\ and\ \citenamefont {Lloyd}}]{Zanardi2004}%
  \BibitemOpen
  \bibfield  {author} {\bibinfo {author} {\bibfnamefont {P.}~\bibnamefont
  {Zanardi}}, \bibinfo {author} {\bibfnamefont {D.~A.}\ \bibnamefont {Lidar}},\
  and\ \bibinfo {author} {\bibfnamefont {S.}~\bibnamefont {Lloyd}},\ }\bibfield
   {title} {\bibinfo {title} {Quantum tensor product structures are observable
  induced},\ }\href {https://doi.org/10.1103/PhysRevLett.92.060402} {\bibfield
  {journal} {\bibinfo  {journal} {Physical Review Letters}\ }\textbf {\bibinfo
  {volume} {92}},\ \bibinfo {pages} {060402} (\bibinfo {year}
  {2004})}\BibitemShut {NoStop}%
\bibitem [{\citenamefont {Schr\"odinger}(1935)}]{schrodinger_1935}%
  \BibitemOpen
  \bibfield  {author} {\bibinfo {author} {\bibfnamefont {E.}~\bibnamefont
  {Schr\"odinger}},\ }\bibfield  {title} {\bibinfo {title} {Discussion of
  probability relations between separated systems},\ }\href
  {https://doi.org/10.1017/S0305004100013554} {\bibfield  {journal} {\bibinfo
  {journal} {Mathematical Proceedings of the Cambridge Philosophical Society}\
  }\textbf {\bibinfo {volume} {31}},\ \bibinfo {pages} {555–563} (\bibinfo
  {year} {1935})}\BibitemShut {NoStop}%
\bibitem [{\citenamefont {Joos}\ and\ \citenamefont {Zeh}(1985)}]{Joos1985}%
  \BibitemOpen
  \bibfield  {author} {\bibinfo {author} {\bibfnamefont {E.}~\bibnamefont
  {Joos}}\ and\ \bibinfo {author} {\bibfnamefont {H.~D.}\ \bibnamefont {Zeh}},\
  }\bibfield  {title} {\bibinfo {title} {The emergence of classical properties
  through interaction with the environment},\ }\href
  {https://doi.org/10.1007/BF01725541} {\bibfield  {journal} {\bibinfo
  {journal} {Zeitschrift für Physik B Condensed Matter}\ }\textbf {\bibinfo
  {volume} {59}},\ \bibinfo {pages} {223} (\bibinfo {year} {1985})}\BibitemShut
  {NoStop}%
\bibitem [{\citenamefont {Zurek}(2003)}]{Zurek2003}%
  \BibitemOpen
  \bibfield  {author} {\bibinfo {author} {\bibfnamefont {W.~H.}\ \bibnamefont
  {Zurek}},\ }\bibfield  {title} {\bibinfo {title} {Decoherence, einselection,
  and the quantum origins of the classical},\ }\href
  {https://doi.org/10.1103/RevModPhys.75.715} {\bibfield  {journal} {\bibinfo
  {journal} {Reviews of Modern Physics}\ }\textbf {\bibinfo {volume} {75}},\
  \bibinfo {pages} {715} (\bibinfo {year} {2003})}\BibitemShut {NoStop}%
\bibitem [{\citenamefont {Schlosshauer}(2005)}]{Schlosshauer2005}%
  \BibitemOpen
  \bibfield  {author} {\bibinfo {author} {\bibfnamefont {M.}~\bibnamefont
  {Schlosshauer}},\ }\bibfield  {title} {\bibinfo {title} {Decoherence, the
  measurement problem, and interpretations of quantum mechanics},\ }\href
  {https://doi.org/10.1103/RevModPhys.76.1267} {\bibfield  {journal} {\bibinfo
  {journal} {Rev. Mod. Phys.}\ }\textbf {\bibinfo {volume} {76}},\ \bibinfo
  {pages} {1267} (\bibinfo {year} {2005})}\BibitemShut {NoStop}%
\bibitem [{\citenamefont {Cotler}\ \emph {et~al.}(2019)\citenamefont {Cotler},
  \citenamefont {Penington},\ and\ \citenamefont {Ranard}}]{Cotler2019}%
  \BibitemOpen
  \bibfield  {author} {\bibinfo {author} {\bibfnamefont {J.~S.}\ \bibnamefont
  {Cotler}}, \bibinfo {author} {\bibfnamefont {G.~R.}\ \bibnamefont
  {Penington}},\ and\ \bibinfo {author} {\bibfnamefont {D.~H.}\ \bibnamefont
  {Ranard}},\ }\bibfield  {title} {\bibinfo {title} {Locality from the
  spectrum},\ }\href {https://doi.org/10.1007/s00220-019-03376-w} {\bibfield
  {journal} {\bibinfo  {journal} {Communications in Mathematical Physics}\
  }\textbf {\bibinfo {volume} {368}},\ \bibinfo {pages} {1267} (\bibinfo {year}
  {2019})}\BibitemShut {NoStop}%
\bibitem [{\citenamefont {Berry}()}]{Berry}%
  \BibitemOpen
  \bibfield  {author} {\bibinfo {author} {\bibfnamefont {M.~V.}\ \bibnamefont
  {Berry}},\ }\bibinfo {title} {Quantum chaology (the bakerian lecture)},\ in\
  \href {https://doi.org/10.1142/9789813221215_0025} {\emph {\bibinfo
  {booktitle} {A Half-Century of Physical Asymptotics and Other Diversions}}},\
  Chap.\ \bibinfo {chapter} {3.3}, pp.\ \bibinfo {pages} {307--322}\BibitemShut
  {NoStop}%
\bibitem [{\citenamefont {Haake}(2001)}]{Haake2001}%
  \BibitemOpen
  \bibfield  {author} {\bibinfo {author} {\bibfnamefont {F.}~\bibnamefont
  {Haake}},\ }\href {https://doi.org/10.1007/978-3-662-04506-0} {\emph
  {\bibinfo {title} {Quantum Signatures of Chaos}}}\ (\bibinfo  {publisher}
  {Springer Berlin Heidelberg},\ \bibinfo {year} {2001})\BibitemShut {NoStop}%
\bibitem [{\citenamefont {Freedman}\ and\ \citenamefont
  {Zini}(2021{\natexlab{a}})}]{Freedman2021a}%
  \BibitemOpen
  \bibfield  {author} {\bibinfo {author} {\bibfnamefont {M.}~\bibnamefont
  {Freedman}}\ and\ \bibinfo {author} {\bibfnamefont {M.~S.}\ \bibnamefont
  {Zini}},\ }\bibfield  {title} {\bibinfo {title} {The universe from a single
  particle},\ }\href {https://doi.org/10.1007/JHEP01(2021)140} {\bibfield
  {journal} {\bibinfo  {journal} {Journal of High Energy Physics}\ }\textbf
  {\bibinfo {volume} {2021}},\ \bibinfo {pages} {140} (\bibinfo {year}
  {2021}{\natexlab{a}})}\BibitemShut {NoStop}%
\bibitem [{\citenamefont {Freedman}\ and\ \citenamefont
  {Zini}(2021{\natexlab{b}})}]{Freedman2021b}%
  \BibitemOpen
  \bibfield  {author} {\bibinfo {author} {\bibfnamefont {M.}~\bibnamefont
  {Freedman}}\ and\ \bibinfo {author} {\bibfnamefont {M.~S.}\ \bibnamefont
  {Zini}},\ }\bibfield  {title} {\bibinfo {title} {The universe from a single
  particle. part ii},\ }\href {https://doi.org/10.1007/JHEP10(2021)102}
  {\bibfield  {journal} {\bibinfo  {journal} {Journal of High Energy Physics}\
  }\textbf {\bibinfo {volume} {2021}},\ \bibinfo {pages} {102} (\bibinfo {year}
  {2021}{\natexlab{b}})}\BibitemShut {NoStop}%
\bibitem [{\citenamefont {Wigner}(1955)}]{Wigner2023}%
  \BibitemOpen
  \bibfield  {author} {\bibinfo {author} {\bibfnamefont {E.~P.}\ \bibnamefont
  {Wigner}},\ }\bibfield  {title} {\bibinfo {title} {Characteristic vectors of
  bordered matrices with infinite dimensions},\ }\href
  {http://www.jstor.org/stable/1970079} {\bibfield  {journal} {\bibinfo
  {journal} {Annals of Mathematics}\ }\textbf {\bibinfo {volume} {62}},\
  \bibinfo {pages} {548} (\bibinfo {year} {1955})}\BibitemShut {NoStop}%
\bibitem [{\citenamefont {Guhr}\ \emph {et~al.}(1998)\citenamefont {Guhr},
  \citenamefont {Müller–Groeling},\ and\ \citenamefont
  {Weidenmüller}}]{Guhr1998}%
  \BibitemOpen
  \bibfield  {author} {\bibinfo {author} {\bibfnamefont {T.}~\bibnamefont
  {Guhr}}, \bibinfo {author} {\bibfnamefont {A.}~\bibnamefont
  {Müller–Groeling}},\ and\ \bibinfo {author} {\bibfnamefont {H.~A.}\
  \bibnamefont {Weidenmüller}},\ }\bibfield  {title} {\bibinfo {title}
  {Random-matrix theories in quantum physics: common concepts},\ }\href
  {https://doi.org/10.1016/S0370-1573(97)00088-4} {\bibfield  {journal}
  {\bibinfo  {journal} {Physics Reports}\ }\textbf {\bibinfo {volume} {299}},\
  \bibinfo {pages} {189} (\bibinfo {year} {1998})}\BibitemShut {NoStop}%
\bibitem [{\citenamefont {Atas}\ \emph {et~al.}(2013)\citenamefont {Atas},
  \citenamefont {Bogomolny}, \citenamefont {Giraud},\ and\ \citenamefont
  {Roux}}]{Atas2013}%
  \BibitemOpen
  \bibfield  {author} {\bibinfo {author} {\bibfnamefont {Y.~Y.}\ \bibnamefont
  {Atas}}, \bibinfo {author} {\bibfnamefont {E.}~\bibnamefont {Bogomolny}},
  \bibinfo {author} {\bibfnamefont {O.}~\bibnamefont {Giraud}},\ and\ \bibinfo
  {author} {\bibfnamefont {G.}~\bibnamefont {Roux}},\ }\bibfield  {title}
  {\bibinfo {title} {Distribution of the ratio of consecutive level spacings in
  random matrix ensembles},\ }\href
  {https://doi.org/10.1103/PhysRevLett.110.084101} {\bibfield  {journal}
  {\bibinfo  {journal} {Phys. Rev. Lett.}\ }\textbf {\bibinfo {volume} {110}},\
  \bibinfo {pages} {84101} (\bibinfo {year} {2013})}\BibitemShut {NoStop}%
\bibitem [{\citenamefont {Broyden}(1970)}]{BROYDEN1970}%
  \BibitemOpen
  \bibfield  {author} {\bibinfo {author} {\bibfnamefont {C.~G.}\ \bibnamefont
  {Broyden}},\ }\bibfield  {title} {\bibinfo {title} {The convergence of a
  class of double-rank minimization algorithms 1. general considerations},\
  }\href {https://doi.org/10.1093/imamat/6.1.76} {\bibfield  {journal}
  {\bibinfo  {journal} {IMA Journal of Applied Mathematics}\ }\textbf {\bibinfo
  {volume} {6}},\ \bibinfo {pages} {76} (\bibinfo {year} {1970})}\BibitemShut
  {NoStop}%
\bibitem [{\citenamefont {Fletcher}(1970)}]{Fletcher1970}%
  \BibitemOpen
  \bibfield  {author} {\bibinfo {author} {\bibfnamefont {R.}~\bibnamefont
  {Fletcher}},\ }\bibfield  {title} {\bibinfo {title} {A new approach to
  variable metric algorithms},\ }\href
  {https://doi.org/10.1093/comjnl/13.3.317} {\bibfield  {journal} {\bibinfo
  {journal} {The Computer Journal}\ }\textbf {\bibinfo {volume} {13}},\
  \bibinfo {pages} {317} (\bibinfo {year} {1970})}\BibitemShut {NoStop}%
\bibitem [{\citenamefont {Goldfarb}(1970)}]{Goldfarb1970}%
  \BibitemOpen
  \bibfield  {author} {\bibinfo {author} {\bibfnamefont {D.}~\bibnamefont
  {Goldfarb}},\ }\bibfield  {title} {\bibinfo {title} {A family of
  variable-metric methods derived by variational means},\ }\href
  {https://doi.org/10.1090/S0025-5718-1970-0258249-6} {\bibfield  {journal}
  {\bibinfo  {journal} {Mathematics of Computation}\ }\textbf {\bibinfo
  {volume} {24}},\ \bibinfo {pages} {23} (\bibinfo {year} {1970})}\BibitemShut
  {NoStop}%
\bibitem [{\citenamefont {Shanno}(1970)}]{Shanno1970}%
  \BibitemOpen
  \bibfield  {author} {\bibinfo {author} {\bibfnamefont {D.~F.}\ \bibnamefont
  {Shanno}},\ }\bibfield  {title} {\bibinfo {title} {Conditioning of
  quasi-newton methods for function minimization},\ }\href
  {https://doi.org/10.1090/S0025-5718-1970-0274029-X} {\bibfield  {journal}
  {\bibinfo  {journal} {Mathematics of Computation}\ }\textbf {\bibinfo
  {volume} {24}},\ \bibinfo {pages} {647} (\bibinfo {year} {1970})}\BibitemShut
  {NoStop}%
\bibitem [{\citenamefont {Fletcher}(2000)}]{Fletcher2000}%
  \BibitemOpen
  \bibfield  {author} {\bibinfo {author} {\bibfnamefont {R.}~\bibnamefont
  {Fletcher}},\ }\href {https://doi.org/10.1002/9781118723203} {\emph {\bibinfo
  {title} {Practical Methods of Optimization}}}\ (\bibinfo  {publisher} {John
  Wiley Sons, Ltd},\ \bibinfo {year} {2000})\BibitemShut {NoStop}%
\bibitem [{\citenamefont {Dumitriu}\ and\ \citenamefont
  {Edelman}(2002)}]{Dumitriu2002}%
  \BibitemOpen
  \bibfield  {author} {\bibinfo {author} {\bibfnamefont {I.}~\bibnamefont
  {Dumitriu}}\ and\ \bibinfo {author} {\bibfnamefont {A.}~\bibnamefont
  {Edelman}},\ }\bibfield  {title} {\bibinfo {title} {Matrix models for beta
  ensembles},\ }\href {https://doi.org/10.1063/1.1507823} {\bibfield  {journal}
  {\bibinfo  {journal} {Journal of Mathematical Physics}\ }\textbf {\bibinfo
  {volume} {43}},\ \bibinfo {pages} {5830} (\bibinfo {year}
  {2002})}\BibitemShut {NoStop}%
\bibitem [{\citenamefont {Mehta}(2004)}]{Mehta2004}%
  \BibitemOpen
  \bibfield  {author} {\bibinfo {author} {\bibfnamefont {M.~L.}\ \bibnamefont
  {Mehta}},\ }\href
  {https://doi.org/https://doi.org/10.1016/S0079-8169(04)80088-6} {\emph
  {\bibinfo {title} {Random Matrices}}},\ edited by\ \bibinfo {editor}
  {\bibfnamefont {M.~L.}\ \bibnamefont {Mehta}},\ \bibinfo {series} {Pure and
  Applied Mathematics}, Vol.\ \bibinfo {volume} {142}\ (\bibinfo  {publisher}
  {Elsevier},\ \bibinfo {year} {2004})\ pp.\ \bibinfo {pages}
  {xiii--xiv}\BibitemShut {NoStop}%
\bibitem [{\citenamefont {\ifmmode~\check{S}\else \v{S}\fi{}untajs}\ \emph
  {et~al.}(2020)\citenamefont {\ifmmode~\check{S}\else \v{S}\fi{}untajs},
  \citenamefont {Bon\ifmmode~\check{c}\else \v{c}\fi{}a}, \citenamefont
  {Prosen},\ and\ \citenamefont {Vidmar}}]{lev2020}%
  \BibitemOpen
  \bibfield  {author} {\bibinfo {author} {\bibfnamefont {J.}~\bibnamefont
  {\ifmmode~\check{S}\else \v{S}\fi{}untajs}}, \bibinfo {author} {\bibfnamefont
  {J.}~\bibnamefont {Bon\ifmmode~\check{c}\else \v{c}\fi{}a}}, \bibinfo
  {author} {\bibfnamefont {T.~c.~v.}\ \bibnamefont {Prosen}},\ and\ \bibinfo
  {author} {\bibfnamefont {L.}~\bibnamefont {Vidmar}},\ }\bibfield  {title}
  {\bibinfo {title} {Quantum chaos challenges many-body localization},\ }\href
  {https://doi.org/10.1103/PhysRevE.102.062144} {\bibfield  {journal} {\bibinfo
   {journal} {Phys. Rev. E}\ }\textbf {\bibinfo {volume} {102}},\ \bibinfo
  {pages} {062144} (\bibinfo {year} {2020})}\BibitemShut {NoStop}%
\bibitem [{\citenamefont {Prakash}\ \emph {et~al.}(2021)\citenamefont
  {Prakash}, \citenamefont {Pixley},\ and\ \citenamefont {Kulkarni}}]{jed2021}%
  \BibitemOpen
  \bibfield  {author} {\bibinfo {author} {\bibfnamefont {A.}~\bibnamefont
  {Prakash}}, \bibinfo {author} {\bibfnamefont {J.~H.}\ \bibnamefont
  {Pixley}},\ and\ \bibinfo {author} {\bibfnamefont {M.}~\bibnamefont
  {Kulkarni}},\ }\bibfield  {title} {\bibinfo {title} {Universal spectral form
  factor for many-body localization},\ }\href
  {https://doi.org/10.1103/PhysRevResearch.3.L012019} {\bibfield  {journal}
  {\bibinfo  {journal} {Phys. Rev. Res.}\ }\textbf {\bibinfo {volume} {3}},\
  \bibinfo {pages} {L012019} (\bibinfo {year} {2021})}\BibitemShut {NoStop}%
\bibitem [{\citenamefont {Cotler}\ \emph {et~al.}(2017)\citenamefont {Cotler},
  \citenamefont {Gur-Ari}, \citenamefont {Hanada}, \citenamefont {Polchinski},
  \citenamefont {Saad}, \citenamefont {Shenker}, \citenamefont {Stanford},
  \citenamefont {Streicher},\ and\ \citenamefont {Tezuka}}]{Cotler2017}%
  \BibitemOpen
  \bibfield  {author} {\bibinfo {author} {\bibfnamefont {J.~S.}\ \bibnamefont
  {Cotler}}, \bibinfo {author} {\bibfnamefont {G.}~\bibnamefont {Gur-Ari}},
  \bibinfo {author} {\bibfnamefont {M.}~\bibnamefont {Hanada}}, \bibinfo
  {author} {\bibfnamefont {J.}~\bibnamefont {Polchinski}}, \bibinfo {author}
  {\bibfnamefont {P.}~\bibnamefont {Saad}}, \bibinfo {author} {\bibfnamefont
  {S.~H.}\ \bibnamefont {Shenker}}, \bibinfo {author} {\bibfnamefont
  {D.}~\bibnamefont {Stanford}}, \bibinfo {author} {\bibfnamefont
  {A.}~\bibnamefont {Streicher}},\ and\ \bibinfo {author} {\bibfnamefont
  {M.}~\bibnamefont {Tezuka}},\ }\bibfield  {title} {\bibinfo {title} {Black
  holes and random matrices},\ }\href {https://doi.org/10.1007/JHEP05(2017)118}
  {\bibfield  {journal} {\bibinfo  {journal} {Journal of High Energy Physics}\
  }\textbf {\bibinfo {volume} {2017}},\ \bibinfo {pages} {118} (\bibinfo {year}
  {2017})}\BibitemShut {NoStop}%
\bibitem [{\citenamefont {Maldacena}\ and\ \citenamefont
  {Stanford}(2016)}]{Maldacena2016}%
  \BibitemOpen
  \bibfield  {author} {\bibinfo {author} {\bibfnamefont {J.}~\bibnamefont
  {Maldacena}}\ and\ \bibinfo {author} {\bibfnamefont {D.}~\bibnamefont
  {Stanford}},\ }\bibfield  {title} {\bibinfo {title} {Remarks on the
  sachdev-ye-kitaev model},\ }\href
  {https://doi.org/10.1103/PhysRevD.94.106002} {\bibfield  {journal} {\bibinfo
  {journal} {Physical Review D}\ }\textbf {\bibinfo {volume} {94}},\ \bibinfo
  {pages} {106002} (\bibinfo {year} {2016})}\BibitemShut {NoStop}%
\bibitem [{\citenamefont {D'Alessio}\ \emph {et~al.}(2016)\citenamefont
  {D'Alessio}, \citenamefont {Kafri}, \citenamefont {Polkovnikov},\ and\
  \citenamefont {Rigol}}]{rigol16review}%
  \BibitemOpen
  \bibfield  {author} {\bibinfo {author} {\bibfnamefont {L.}~\bibnamefont
  {D'Alessio}}, \bibinfo {author} {\bibfnamefont {Y.}~\bibnamefont {Kafri}},
  \bibinfo {author} {\bibfnamefont {A.}~\bibnamefont {Polkovnikov}},\ and\
  \bibinfo {author} {\bibfnamefont {M.}~\bibnamefont {Rigol}},\ }\bibfield
  {title} {\bibinfo {title} {From quantum chaos and eigenstate thermalization
  to statistical mechanics and thermodynamics},\ }\href
  {https://doi.org/10.1080/00018732.2016.1198134} {\bibfield  {journal}
  {\bibinfo  {journal} {Advances in Physics}\ }\textbf {\bibinfo {volume}
  {65}},\ \bibinfo {pages} {239} (\bibinfo {year} {2016})},\ \Eprint
  {https://arxiv.org/abs/https://doi.org/10.1080/00018732.2016.1198134}
  {https://doi.org/10.1080/00018732.2016.1198134} \BibitemShut {NoStop}%
\bibitem [{\citenamefont {Bell}\ and\ \citenamefont {Dean}(1970)}]{Bell1970}%
  \BibitemOpen
  \bibfield  {author} {\bibinfo {author} {\bibfnamefont {R.~J.}\ \bibnamefont
  {Bell}}\ and\ \bibinfo {author} {\bibfnamefont {P.}~\bibnamefont {Dean}},\
  }\bibfield  {title} {\bibinfo {title} {Atomic vibrations in vitreous
  silica},\ }\href {https://doi.org/10.1039/DF9705000055} {\bibfield  {journal}
  {\bibinfo  {journal} {Discuss. Faraday Soc.}\ }\textbf {\bibinfo {volume}
  {50}},\ \bibinfo {pages} {55} (\bibinfo {year} {1970})}\BibitemShut {NoStop}%
\bibitem [{\citenamefont {Machta}\ \emph {et~al.}(2013)\citenamefont {Machta},
  \citenamefont {Chachra}, \citenamefont {Transtrum},\ and\ \citenamefont
  {Sethna}}]{sethna13}%
  \BibitemOpen
  \bibfield  {author} {\bibinfo {author} {\bibfnamefont {B.~B.}\ \bibnamefont
  {Machta}}, \bibinfo {author} {\bibfnamefont {R.}~\bibnamefont {Chachra}},
  \bibinfo {author} {\bibfnamefont {M.~K.}\ \bibnamefont {Transtrum}},\ and\
  \bibinfo {author} {\bibfnamefont {J.~P.}\ \bibnamefont {Sethna}},\ }\bibfield
   {title} {\bibinfo {title} {Parameter space compression underlies emergent
  theories and predictive models},\ }\href
  {https://doi.org/10.1126/science.1238723} {\bibfield  {journal} {\bibinfo
  {journal} {Science}\ }\textbf {\bibinfo {volume} {342}},\ \bibinfo {pages}
  {604} (\bibinfo {year} {2013})},\ \Eprint
  {https://arxiv.org/abs/https://www.science.org/doi/pdf/10.1126/science.1238723}
  {https://www.science.org/doi/pdf/10.1126/science.1238723} \BibitemShut
  {NoStop}%
\bibitem [{\citenamefont {Cao}\ \emph {et~al.}(2017)\citenamefont {Cao},
  \citenamefont {Carroll},\ and\ \citenamefont {Michalakis}}]{Cao2017}%
  \BibitemOpen
  \bibfield  {author} {\bibinfo {author} {\bibfnamefont {C.}~\bibnamefont
  {Cao}}, \bibinfo {author} {\bibfnamefont {S.~M.}\ \bibnamefont {Carroll}},\
  and\ \bibinfo {author} {\bibfnamefont {S.}~\bibnamefont {Michalakis}},\
  }\bibfield  {title} {\bibinfo {title} {Space from hilbert space: Recovering
  geometry from bulk entanglement},\ }\href
  {https://doi.org/10.1103/PhysRevD.95.024031} {\bibfield  {journal} {\bibinfo
  {journal} {Phys. Rev. D}\ }\textbf {\bibinfo {volume} {95}},\ \bibinfo
  {pages} {24031} (\bibinfo {year} {2017})}\BibitemShut {NoStop}%
\bibitem [{\citenamefont {Carroll}\ and\ \citenamefont
  {Singh}(2021)}]{Carroll2021}%
  \BibitemOpen
  \bibfield  {author} {\bibinfo {author} {\bibfnamefont {S.~M.}\ \bibnamefont
  {Carroll}}\ and\ \bibinfo {author} {\bibfnamefont {A.}~\bibnamefont
  {Singh}},\ }\bibfield  {title} {\bibinfo {title} {Quantum mereology:
  Factorizing hilbert space into subsystems with quasiclassical dynamics},\
  }\href {https://doi.org/10.1103/PhysRevA.103.022213} {\bibfield  {journal}
  {\bibinfo  {journal} {Physical Review A}\ }\textbf {\bibinfo {volume}
  {103}},\ \bibinfo {pages} {022213} (\bibinfo {year} {2021})}\BibitemShut
  {NoStop}%
\bibitem [{\citenamefont {Bekenstein}(1973)}]{Bekenstein1973}%
  \BibitemOpen
  \bibfield  {author} {\bibinfo {author} {\bibfnamefont {J.~D.}\ \bibnamefont
  {Bekenstein}},\ }\bibfield  {title} {\bibinfo {title} {Black holes and
  entropy},\ }\href {https://doi.org/10.1103/PhysRevD.7.2333} {\bibfield
  {journal} {\bibinfo  {journal} {Physical Review D}\ }\textbf {\bibinfo
  {volume} {7}},\ \bibinfo {pages} {2333} (\bibinfo {year} {1973})}\BibitemShut
  {NoStop}%
\end{thebibliography}%

\newpage

\onecolumngrid

\newpage

{\centering

{\normalsize \bf Supplementary Material: Unveiling Order from Chaos by approximate 2-localization of random matrices }
% {\Large \bf Supplemental Materials: Unsupervised detection of off-diagonal order from diagonal basis measurements}

\vspace{0.5cm}

Nicolas Loizeau$^1$, Flaviano Morone$^{1}$, Dries Sels$^{1, 2}$\\

\vspace{0.1in}

{\it
$^1$Department of Physics, New York University, New York, New York 10003, USA\\
$^2$Center for Computational Quantum Physics, Flatiron Institute,\\
162 Fifth Avenue, New York, New York 10010 USA\\
}}

\vspace{1cm}

\onecolumngrid

\renewcommand{\thesection}{S\arabic{section}}  
\renewcommand{\thefigure}{S\arabic{figure}}

\setcounter{equation}{0}
\setcounter{figure}{0}
\section{SECTION 1 -- Schriefer-Wolff localization}
\label{sup:schrieferwolf}

In practice we localize a spectrum $E_i$ by minimizing the cost $C=\frac{1}{2^N}\sum_i (E_i-\mathcal{E}_i)^2$ where $\mathcal{E}_i$ are the eigenvalues of a local Hamiltonian $H'=\sum h_i \tau_i$. Here we describe an alternative noteworthy procedure that use the Schrieffer-Wolff transformation to find a unitary $U$ that transforms a Hamiltonian $H$ into a local one $H'=U^\dag H U$.

 Any given Hamiltonian can be split into a part that lies in the desired subspace and a part that does not, i.e. $H=H_k+H_\perp$ where $H_k$ denotes the projection of the Hamiltonian on the subspace of $k$-local operators. Once we have such a decomposition we can try to perturbatively remove the undesired part. Let’s write the unitary $U=e^{-S}$ that transform the Hamiltonian as $H'=U^\dag H U=e^{-S}(H_k+H_\perp)e^S$ assuming S is small we expand the above expression in a Taylor series. To lowest order we find
\begin{equation}
    H'=H_k+H_\perp-[S,H_k]-[S,H_\perp]+O(S^2)
\end{equation}
Consequently, to lowest order we can remove the undesired part choosing S such that it removes $H_\perp$. There is one subtlety, namely that $[S, H] = H_\perp$ might not have a solution if $H_\perp$ has a component that is diagonal in $H$. This can
be resolved by only using $H_k$ in the commutator: $[S,H_k]=H_\perp$. This is essentially equivalent to a perturbative Schriefer-Wolff transformation where $H_\perp$ is assumed to be a small
perturbation to $H_k$. The latter can be solved by diagonalizing $H_k=\sum_n\epsilon_n\oprod{n}{n}$, such that the matrix elements of $S$ are:
\begin{equation}
    \bra{n}S\ket{m}=-\frac{\bra{n}H_\perp\ket{m}}{\epsilon_n-\epsilon_m}
\end{equation}
when $n\neq m$ and zero otherwise.
We iterate this procedure until $H_\perp$ is small enough. Each iteration gives a unitary $U_i$ that bring the current $H$ closer to the local form. The final unitary that localizes the initial $H$ is the product $\prod_i U_i$.
Note that if $H_\perp$ is diagonal in the $\ket{n}$ basis, ie if $[H_k, H_\perp]=0$ then $S=0$. In reality $H_\perp$ is not small as compared to $H_k$, unless we are almost converged of course. Consequently we only want to remove a small portion $\alpha$ in every step and we take $U=e^{\alpha S}$. In practice $\alpha=0.1$ works well for systems up to $N = 12$. This also guaranties that once the procedure has converged i.e $U_i=\id$, $S=0$ then the remaining non local part commutes with the local one. Therefore, in addition to giving a basis in which $H$ is local when $H$ is localizable, this procedure gives us a decomposition into a local part that commutes with a non-local part when $H$ is not localizable.

\begin{algorithm}[H]\label{alg:sw}
\caption{Schrieffer-Wolff-based localization algorithm}
\begin{algorithmic}
\State $d \gets$ Dim(H)
\For{iteration $i$}
\State {$H_2 \gets \mathcal{P}_{1,2}(H)$}
\State{$H_\perp \gets H-H_2$}
\State{$\{\epsilon_n\},V \gets$ Diagonalize($H_2$)}
\State{$S\gets -V^\dag H_\perp V$}
\For{$0\leq n<d$, $0\leq m<d$}
\State{$S[n,m] \gets \frac{S[n,m]}{\epsilon_n-\epsilon_m}$}
\EndFor
\For{$0\leq n<d$}
\State{$S[n,n]\gets 0$}
\EndFor
\State{$S\gets VSV^\dag$}
\State{$U_i \gets e^{-S}$}
\State{$H\gets USU^\dag$}
\State{$U \gets U_iU$}
\EndFor
\State \Return {$U$}
\end{algorithmic}
\end{algorithm}

\section{SECTION 2 -- Tighter bound on the rank of a 2-local projector in the N=3 case}
\label{sup:tighterbound}
We showed that any 2-localizable projector of size $2^N$ has rank greater than $\frac{2^N}{M}$.
It is possible to derive a tighter bound by considering the quantity $\bra{q}\tau_i\ket{q}$. Using Cauchy-Schwartz inequality we have 
\begin{align}
K = \frac{1}{2^N}\sum_i\left(\sum_q \bra{q}\tau_i\ket{q}\right)^2 \leq \frac{K}{2^N}\sum_i\sum_q \bra{q}\tau_i\ket{q}^2.
\end{align}
Now, note that the sum $\sum_i \bra{q}\tau_i\ket{q}^2$ can be decomposed into two sums: one over the 1-local strings and the other on the 2-local-only strings.
Let's examine the quantity $\bra{q}\tau_i\ket{q}$ in the case that $\tau_i$ is 1-local. For simplicity we take $N=3$. A particular 1-local $\tau$ can be written as $\tau=\sigma_{i}^A\otimes \id^B\otimes \id^C$. Define $\rho_q= \oprod{q}{q}$. Then $\bra{q}\tau\ket{q}=\Tr(\rho_q\tau)=\Tr(\rho^A_q \sigma_i)$ where $\rho^A=\frac{\id +\overrightarrow{\alpha_A}\cdot\overrightarrow{\sigma}}{2}$ is the reduced density matrix on subsystem $A$. Now given that $\Tr(\rho^A_q \sigma_i)=\alpha_{Ai}$ and that $|\overrightarrow{\alpha_A}|\leq 1$, we can see that 
\begin{align}
    \sum_{\tau_i \in 1-loc} \bra{q}\tau_i\ket{q}^2=|\overrightarrow{\alpha_A}|^2+|\overrightarrow{\alpha_B}|^2+|\overrightarrow{\alpha_C}|^2 \leq 3
\end{align}
We can use the same idea to derive a bound on the second sum: take $\tau$ to be 2-local-only string. Then $\bra{q}\tau\ket{q}=tr(\sigma_i^A\otimes\sigma_j^B \rho_{AB})$ with $\rho_{AB}=\frac{1}{2}(\id_{AB}+\overrightarrow{\alpha_A}\cdot\overrightarrow{\sigma}\otimes\id_B +\id_A \otimes \overrightarrow{\alpha_B}\cdot\overrightarrow{\sigma}+\sum_{k,l}\beta_{AB\ k,l}\sigma_k^A \otimes \sigma_l^B)$. When computing the trace of $\sigma_i^A\otimes\sigma_j^B \rho_{AB}$, the one-body terms in $\alpha$ and $\beta$ have zero contribution and $\bra{q}\tau\ket{q}=\frac{1}{4}tr(\sum_{k,l}\beta_{AB\ k,l}\sigma_i^A\sigma_k^A \otimes \sigma_j^B\sigma_l^B)=\beta_{AB\ i,j}$. We eventually have
\begin{align}
    \sum_{\tau_i \in 2-loc-only} \bra{q}\tau_i\ket{q}^2=|\overrightarrow{\beta_{AB}}|^2+|\overrightarrow{\beta_{BC}}|^2+|\overrightarrow{\beta_{AC}}|^2 \leq 3
\end{align}
and 
\begin{align}
    \sum_i \bra{q}\tau_i\ket{q}^2 \leq 6
\end{align}
this yields $K\geq \frac{4}{3}$ in the N=3 case.

\section{SECTION 3 -- Gradient and Hessian derivations}
\label{sup:gradienthessian}
Here we derive the gradient and the hessian of the cost function \begin{equation}
    C(h) = \frac{1}{2^N}\sum_n(E_n-\Eps_n)^2
\end{equation} where $\mathcal{E}_n$ are the eigenvalues of a local hamiltonian $H'=\sum_i h_i \tau_i$. Let's compute the gradient and the hessian of $C$ with respect to $h$.
Fisrt taking the derivative of $H'\ket{n}=\Eps_n\ket{n}$ with respect to $h_i$: $(\partial_iH')\ket{n}+H'\partial_i\ket{n}=\partial_i(\Eps_n)\ket{n}+\Eps_n\partial_i\ket{n}$, multiplying by $\bra{n}$ to the left we get $\bra{n}\tau_i\ket{n}+\bra{n}\Eps_n\partial_i\ket{n}=\partial_i(\Eps_n)+\Eps_n\bra{n}\partial_i\ket{n}\implies$\begin{equation}
    \partial_i \Eps_n=\bra{n}\tau_i\ket{n}.
    \label{eq:diE}
\end{equation}
Now, $C=\frac{1}{2^{N+1}}\left(\sum_n E_n^2+\sum_n \Eps_n^2-2\sum_nE_n\Eps_n\right)=\frac{1}{2^{N+1}}\left(\sum_n |H|^2+2^N\sum_i h_i^2-2\sum_nE_n\Eps_n\right)$ and from eq \ref{eq:diE} we have \begin{equation}
    \partial_iC=h_i-\frac{1}{2^N}\sum_n E_n\bra{n}\tau_i\ket{n}.
    \label{eq:gradient}
\end{equation}
To compute the Hessian we first go back to $\tau_i\ket{n}+H'\partial_i\ket{n}=\partial_i(\Eps_n)\ket{n}+\Eps_n\partial_i\ket{n}$ but this time we multiply by $\bra{m}$ to the left. This gives \begin{equation}
    \bra{m}\tau_i\ket{n}=(\Eps_n-\Eps_m)\bra{m}\partial_i\ket{n}.
    \label{eq:mdin}
\end{equation}
Now, differentiate the gradient : $\partial_i\partial_j C =\delta_{ij}-\frac{1}{2^N}\sum_n E_n\left((\partial_i\bra{n})\tau_j\ket{n}+\bra{n}\tau_j\partial_i\ket{n}\right)$. Inserting $\sum_m\oprod{m}{m}:\partial_i\partial_j C =\delta_{ij}-\frac{1}{2^N}\sum_n\sum_m E_n\left((\partial_i\bra{n})\oprod{m}{m}\tau_j\ket{n}+\bra{n}\tau_j\oprod{m}{m}\partial_i\ket{n}\right)$. Now separating $n=m$ and $n\neq m$ and using eq \ref{eq:mdin} in the $n\neq m$ case:
\begin{align}
    g_{ij}=\partial_i\partial_j C=\delta_{ij}&-\frac{1}{2^N}\sum_n\sum_{m\neq n} E_n\left(\frac{\bra{n}\tau_i\ket{m}}{\Eps_n-\Eps_m}\bra{m}\tau_j\ket{n}+\bra{n}\tau_j\ket{m}\frac{\bra{m}\tau_i\ket{n}}{\Eps_n-\Eps_m}
    \right)\\
    &-\frac{1}{2^N}\sum_n E_n\left((\partial_i\bra{n})\oprod{n}{n}\tau_j\ket{n}+\bra{n}\tau_j\oprod{n}{n}\partial_i\ket{n}\right)
\end{align}
but note that $\partial_i\iprod{n}{n}=0\implies (\partial_i\bra{n})\ket{n}=-\bra{n}\partial_i\ket{n}$ so the second sum is $0$ and we have
\begin{align}
    g_{ij}=\delta_{ij}-\frac{1}{2^N}\sum_n\sum_{m\neq n} \frac{E_n}{\Eps_n-\Eps_m}\big(\bra{n}\tau_i\ket{m}\bra{m}\tau_j\ket{n}+\bra{n}\tau_j\ket{m}\bra{m}\tau_i\ket{n}
    \big)
\end{align}

Now let's compute $\partial_i\partial_j C$ in the minimum i.e when $E_i=\Eps_i$. First, remark that $\partial_i\partial_j C=\delta_{ij}-\frac{1}{2^N}\sum_n\sum_{m\neq n} \frac{E_n}{\Eps_n-\Eps_m}T^{nm}_{ij}$ where $T^{nm}_{ij}$ is symmetric in $n,m$. Moreover, \begin{align}
    \frac{1}{2}\sum_{n, m\neq n}\frac{E_n-E_m}{\Eps_n-\Eps_m}T^{nm}_{ij}&=\frac{1}{2}\sum_{n,m\neq n}\frac{E_n}{\Eps_n-\Eps_m}T^{nm}_{ij}-\frac{1}{2}\sum_{n, m\neq n}\frac{E_m}{\Eps_n-\Eps_m}T^{nm}_{ij}\\&=\sum_{n, m\neq n}\frac{E_n}{\Eps_n-\Eps_m}T^{nm}_{ij}
\end{align}
In the minimum, $\frac{E_n-E_m}{\Eps_n-\Eps_m}=1$ for $n\neq m$ and 
\begin{align}
    g_{ij}^0&=\delta_{ij}-\frac{1}{2^{N+1}}\sum_n\sum_{m\neq n} \bra{n}\tau_i\ket{m}\bra{m}\tau_j\ket{n}+\bra{n}\tau_j\ket{m}\bra{m}\tau_i\ket{n}\nonumber\\
    &= \delta_{ij}-\frac{1}{2^{N+1}}\sum_n\sum_{m} \bra{n}\tau_i\ket{m}\bra{m}\tau_j\ket{n}+\bra{n}\tau_j\ket{m}\bra{m}\tau_i\ket{n}+\frac{1}{2^{N+1}}\sum_{n}\bra{n}\tau_i\ket{n}\bra{n}\tau_j\ket{n}+\bra{n}\tau_j\ket{n}\bra{n}\tau_i\ket{n}\nonumber\\
    &=\frac{1}{2^N}\sum_{n}\bra{n}\tau_i\ket{n}\bra{n}\tau_j\ket{n}
\end{align}
Note that in the minimum, $h^0$ is an eigenvector of $g^0$ with eigenvalue 1. $    \sum_i g_{ij}h_i=\frac{1}{2^N}\sum_{n,i}\bra{n}\tau_i\oprod{n}{n}\tau_j\ket{n}h_i=\frac{1}{2^N}\sum_{n}\bra{n}H'\oprod{n}{n}\tau_j\ket{n}=\frac{1}{2^N}\sum_{n,i}\Eps_n\bra{n}\tau_j\ket{n}$. In the minimum, $\partial_i C=0$ hence, using eq \ref{eq:gradient}, we have \begin{equation}
    \sum_i g_{ij}^0h_i^0=h_j^0.
\end{equation}
In the case the $\tau_i$ are diagonal, 
\begin{align}
    g_{ij}^0&=\frac{1}{2^N}\sum_{n}\tau_{i,nn}\tau_{j,nn}\\
    &=\frac{1}{2^N}\sum_{n}\Tr(\tau_{i}\tau_{j})\\
    &=\delta_{ij}.
\end{align} 

\section{SECTION 4 -- Hessian eigenvalues}
\label{sup:hessianev}
Let us derive a general formula for the eigenvalues of $g^0$ the Hessian in the minimum.
Starting from 
\begin{equation}
    g_{ij}^0=\frac{1}{2^N}\sum_{n}\bra{n}\tau_i\ket{n}\bra{n}\tau_j\ket{n},
\end{equation} 
define $Q_{ni}=\frac{1}{\sqrt{2^N}}\bra{n}\tau_i\ket{n}$ and note that $g_{ij}^0=Q^TQ$. Consider the singular vectors $v_k,f_k$ and singular values $\mu_k$ of $Q$: $Qv_k=\mu_kf_k$ and $Q^Tf_k=\mu_kv_k$. Multiplying these eigenvalue equations by $Q$ or $Q^T$ we get $g^0v_k=\mu_k^2v_k$ and $g^0f_k=\mu_k^2f_k$ i.e $\lambda_k = \mu_k^2$ is an eigenvalue of $g^0$ .

Now, remark that $f_k^TQQ^Tf_k=f_k^T\mu^2f_k$ hence we have \begin{equation}
    \lambda_k = \mu_k^2 = \frac{f_k^TQQ^Tf_k}{f_k^Tf_k}.
\end{equation}
Using the definition of $Q$, we can rewrite the numerator :
\begin{align}
    f_k^TQQ^Tf_k&=\frac{1}{2^N}\sum_i\sum_{n,m} f_n\bra{n}\tau_i\ket{n}f_m\bra{m}\tau_i\ket{m}\\
    &=\frac{1}{2^N}\sum_iTr\left(\sum_{n} f_n\oprod{n}{n}\tau_i\right)Tr\left(\sum_{m} f_m\oprod{m}{m}\tau_i\right)
\end{align}
defining $F(H)=\sum_{m} f_m\oprod{m}{m}\tau_i$ we have $f_k^TQQ^Tf_k=\frac{1}{2^N}Tr(F(H)\tau_i)^2$.
Similarly, for the denominator, $f_k^Tf_k=\sum_nf_{kn}^2=Tr\left(\sum_{n,m}f_n\oprod{n}{n}f_m\oprod{m}{m}\right)=Tr(F(H)^2)$ hence, the eigenvalues of $g^0$ can be written as
\begin{equation}
    \lambda_k=\frac{1}{2^N} \frac{\sum_iTr(F(H)\tau_i)^2}{Tr(F(H)^2)}.
    \label{lambda_k}
\end{equation}

\section{SECTION 5 -- Computing the Hessian's second largest eigenvalue}
\label{sup:hessian2ev}
Here we derive a formula for $\lambda_2$ assuming that $F_2(H)$ is the traceless part of $H^2$ and that the $\tau_i$ are diagonal 2-local Pauli strings. Note that this is not equivalent to computing the second eigenvalue of $g^0$ in the diagonal case. In the diagonal case all the eigenvalues of $g^0$ are $1$. Instead we are using equation \eqref{lambda_k} with a set of diagonal $\tau_i$ in order to get an estimate of $\lambda_2$ in the general case.
Start with
\begin{equation}
    \lambda_2 = \frac{1}{2^N}\frac{\sum_i\Tr\left(F_2(H)\tau_i\right)^2}{\Tr\left(F_2(H)^2\right)}
    \label{lambda_2}
\end{equation}
where $F_2(H)$ is the traceless part of $H^2$ i.e $F_2(H)=\sum_{i\neq j}h_ih_j\tau_i \tau_j$.
Define numerator $\alpha=\sum_i\Tr\left(F_2(H)\tau_i\right)^2$ and denominator $\beta=\Tr\left(F_2(H)^2\right)$.

\subsection{Numerator}
First we focus on the numerator 
\begin{align}
    \alpha &=\sum_k\left(\sum_{i\neq j}h_ih_j\Tr(\tau_i\tau_j\tau_k)\right)^2
    \label{numerator}.
\end{align}

Check that the constraint $i\neq j$ is irrelevant:
\begin{align}
    \sum_{i\neq j}h_ih_j\Tr(\tau_i\tau_j\tau_k)&=\sum_{i, j}h_ih_j\Tr(\tau_i\tau_j\tau_k)-\sum_{i}h_i^2\Tr(\tau_i\tau_i\tau_k)\\
    &=\sum_{i, j}h_ih_j\Tr(\tau_i\tau_j\tau_k)
\end{align}
because $\Tr(\tau_i\tau_i\tau_k)=\Tr(\tau_k)=0$.

Now relabel the indices of the Pauli strings into the indices of the couplings: $i\rightarrow(a>b)$, $j\rightarrow(c>d)$, $k\rightarrow(e>f)$ and define the notation $\Tr(ab,cd,ef)=\Tr(Z_a\otimes Z_b\cdot Z_c\otimes Z_d\cdot Z_e\otimes Z_f)$
First consider the part inside the parenthesis in eq \eqref{numerator}. 

\begin{align}
\sum_{a>b}\sum_{c>d}J_{ab}J_{cd}\Tr(ab,cd,ef)&=\sum_{a>b}\frac{1}{2}\left[\sum_{cd}J_{ab}J_{cd}\Tr(ab,cd,ef)-\sum_{c}J_{ab}J_{cc}\Tr(ab,cc,ef)\right]
\end{align}
because $J_{cc}=0$, we have 
\begin{align}
\sum_{a>b}\sum_{c>d}J_{ab}J_{cd}\Tr(ab,cd,ef)&=\sum_{a>b}\frac{1}{2}\left[\sum_{cd}J_{ab}J_{cd}\Tr(ab,cd,ef)\right]
\end{align}
and similarly 
\begin{align}
\sum_{a>b}\sum_{c>d}J_{ab}J_{cd}\Tr(ab,cd,ef)&=\frac{1}{4}\sum_{ab}\sum_{cd}J_{ab}J_{cd}\Tr(ab,cd,ef)
\end{align}
Now,
\begin{align}
    \alpha &= \sum_{e>f}\left[\frac{1}{4}\sum_{ab}\sum_{cd}J_{ab}J_{cd}\Tr(ab,cd,ef)\right]^2\\
    &=\frac{1}{16}\frac{1}{2}\left(\sum_{ef}\left[\sum_{abcd}J_{ab}J_{cd}\Tr(ab,cd,ef)\right]^2-\sum_{e}\left[\sum_{abcd}J_{ab}J_{cd}\Tr(ab,cd,ee)\right]^2\right)
\end{align}
And define 
\begin{equation}
    \alpha'= \frac{1}{32}\sum_{e}\left[\sum_{abcd}J_{ab}J_{cd}\Tr(ab,cd,ee)\right]^2
\end{equation}
so that 
\begin{equation}
    \alpha = \frac{1}{32}\sum_{ef}\left[\sum_{abcd}J_{ab}J_{cd}\Tr(ab,cd,ef)\right]^2-\alpha'
\end{equation}

Now let's split the $e,f$ sum in $e\neq f$ and $e$. This cancels the $\alpha'$ term:

\begin{align}
    \alpha &= \frac{1}{32}\sum_{e\neq f}\left[\sum_{abcd}J_{ab}J_{cd}\Tr(ab,cd,ef)\right]^2+\frac{1}{32}\sum_{e}\left[\sum_{abcd}J_{ab}J_{cd}\Tr(ab,cd,ee)\right]^2-\alpha'\\
    &=\frac{1}{32}\sum_{e\neq f}\left[\sum_{abcd}J_{ab}J_{cd}\Tr(ab,cd,ef)\right]^2
\end{align}
The trace $\Tr(ab,cd,ef)$ has only two possibles values : $0$ or $2^N$. Computing $\alpha$ is therefore a matter of figuring out what combinations of $a,b,c,d,e,f$ have non zero trace. $\Tr(ab,cd,ef)=\Tr(Z_a\otimes Z_b\cdot Z_c\otimes Z_d\cdot Z_e\otimes Z_f)$ is non zero when there is a even number of $Z$ on each site. This suggests a diagrammatic method for computing $\Tr(ab,cd,ef)$ where each non zero trace term in the sum is represented by a graph. Consider a graph where the nodes are the indices $a,b,c,d,e,f$ and an edge between two nodes means that the two indices are equal.
A simple set of rules give non zero diagrams: 
\begin{itemize}
    \item edges $ab,cd,ef$ are forbidden by $J_{aa}=0$ and $e\neq f$
    \item connected components have an even number of nodes
    \item edges $ab,cd,ef$ are forbidden by transitivity e.g the couple of edges $ac,cb$ is forbidden.
\end{itemize}
There are 8 contributing diagrams:
\begin{equation}
\diagramt{{ac,be,df}}{a}\quad\quad
\diagramt{{af,bd,ce}}{a}\quad\quad
\diagramt{{af,bc,de}}{a}\quad\quad
\diagramt{{ad,be,cf}}{a}
\end{equation}
\begin{equation}
\diagramt{{ae,bc,df}}{a}\quad\quad
\diagramt{{ad,bf,ce}}{a}\quad\quad
\diagramt{{ae,cf,bd}}{a}\quad\quad
\diagramt{{ac,bf,de}}{a}
\label{diagrams3}
\end{equation}
Consider the first one (all 8 diagrams have the same contribution)
\begin{align}
    \alpha&=\frac{1}{32}\sum_{e\neq f}\left[\sum_{abcd}J_{ab}J_{cd}\Tr(ab,cd,ef)\right]^2\\
    &=\frac{1}{32}\sum_{e\neq f}\left[8\sum_{abcd}J_{ab}J_{cd}2^N\delta_{ac}\delta_{be}\delta_{df}\right]^2\\
    &=\frac{64}{32}4^N\sum_{e\neq f}\left[\sum_{a}J_{ae}J_{af}\right]^2\\
    &=2\cdot 4^N\sum_{ef}\left[\sum_{a}J_{ae}J_{af}\right]^2-2\cdot 4^N\sum_{e}\left[\sum_{a}J_{ae}J_{ea}\right]^2\\
    &=2\cdot 4^N\sum_{efab}J_{ae}J_{af}J_{be}J_{bf}-2\cdot 4^N\sum_{e}\left[(J^2)_{ee}\right]^2\\
    &=2\cdot 4^N\left(\Tr(J^4)-\sum_{e}\left[(J^2)_{ee}\right]^2\right)
\end{align}

\subsection{Denominator}

\begin{align}    
    \beta &=\Tr\left(\left[\sum_{i\neq j}h_ih_j\tau_i\tau_j\right]^2\right)\\
    &=\Tr\left(\left[\sum_{ij}h_ih_j\tau_i\tau_j-\sum_{i}h_i^2\id\right]^2\right)\\
    &=\Tr\left( \sum_{ijkl}h_ih_jh_kh_l\tau_i\tau_j\tau_k\tau_l+\left[\sum_{i}h_i^2 \right]^2\id- 2\left[\sum_{i}h_i^2\id\right]\sum_{ij}h_ih_j\tau_i\tau_j \right)\\
    &=\sum_{ijkl}h_ih_jh_kh_l\Tr(\tau_i\tau_j\tau_k\tau_l)+2^N\left[\sum_{i}h_i^2\right]^2-2\left[\sum_{i}h_i^2\right]\sum_{ij}h_ih_j\Tr(\tau_i\tau_j)\\
    &=\sum_{ijkl}h_ih_jh_kh_l\Tr(\tau_i\tau_j\tau_k\tau_l)+2^N\left[\sum_{i}h_i^2\right]^2-2\left[\sum_{i}h_i^2\right]\sum_{ij}h_ih_j\delta_{ij}2^N\\
    &=\sum_{ijkl}h_ih_jh_kh_l\Tr(\tau_i\tau_j\tau_k\tau_l)-2^N\left[\sum_{i}h_i^2\right]^2\\
    &=\beta'\quad\quad\quad\quad\quad\quad \quad\quad\quad\quad  +\quad \beta''
\end{align}

The second term is
\begin{align}
    \beta'' &= -2^N\left[\sum_{i}h_i^2\right]^2\\
    &=-2^N\left[\sum_{a>b}J_{ab}^2\right]^2\\
    &=-2^N\left[\frac{1}{2}\sum_{ab}J_{ab}^2\right]^2\\
    &=-\frac{2^N}{4}\left[\Tr J^2\right]^2
\end{align}

Now consider $\beta'=\sum_{ijkl}h_ih_jh_kh_l\Tr(\tau_i\tau_j\tau_k\tau_l)$ and relabel it using the sites and the couplings. The condition $a>b$ brings a $\frac{1}{2}$ for each sum because all the diagonals terms are zero since $J_{aa}=0$.
\begin{align}
    \beta' &= \sum_{ijkl}h_ih_jh_kh_l\Tr(\tau_i\tau_j\tau_k\tau_l)\\
    &=\sum_{a>b}\sum_{c>d}\sum_{e>f}\sum_{g>h}J_{ab}J_{cd}J_{ef}J_{gh}\Tr(ab,cd,ef,gh)\\
    &=\frac{1}{2^4}\sum_{ab}\sum_{cd}\sum_{ef}\sum_{gh}J_{ab}J_{cd}J_{ef}J_{gh}\Tr(ab,cd,ef,gh)
\end{align}
We can now compute this sum using diagrams.
\subsubsection*{Diagram inclusion}
Here the use of 4 lines diagrams is tricky because some diagrams include other ones. 
For example 
\begin{align}
    \diagramf{{ac,ce,eg,bd,fh}}{a}\subset \diagramf{{ac,eg,bd,fh}}{a}
\end{align}
because all the constraints represented in the second diagram are included in the constraints represented by the first diagram hence all the terms represented by the first diagram are included in the second diagram's terms.
It is not enough to only consider higher order diagram (diagram with less constraints) because some diagram overlap i.e share some terms.
\subsubsection*{Diagram overlap}
Some diagrams overlap e.g:
\begin{equation}
    \diagramf{{ac,eg,bd,fh}}{a} \cap \diagramf{{ae,cg,bd,fh}}{a} = \diagramf{{ac,ce,eg,bd,fh}}{a}
\end{equation}
Hence, even if we only sum over the higher order diagrams, some contributions are counted twice so we need to subtract them once. In this example, the third diagram needs to be subtracted when summing the terms of the two first diagrams.

\subsubsection*{Diagram counting} 

In order to solve this diagram overlap problem, we count every individual diagram (not only the high order ones), but every time we count one diagram, we subtract all the included diagrams. Doing so, we consider that 2 disconnected components of one diagram are explicitly different and hence if we allow then to be equal in the sum, we need to subtract the included diagrams where they are equal.

Consider the chain
\begin{align}
     \diagramf{{ac,eg,bd,fh}}{a}\supset\diagramf{{ac,ce,eg,bd,fh}}{a}\supset\diagramf{{ac,ce,df,eg,bd,fh}}{a}
\end{align}

When counting the first diagram, the second one has to be subtracted from it. But the third one has to be subtracted from the second one. So the third one has to be added to the first one.
In general, the weight of a diagram can be recursivelly assigned by doing a DFS through the graph of inclusion of the diagrams as in the following algorithm:

\begin{algorithm}[H]\label{alg:sw}
\caption{Assigning weights to each diagram}
\begin{algorithmic}
\Procedure{Count}{diagram D, integer v}
\State {D.weight += v}
\For{each diagram E included in D}
\State {Count(E,-v)}
\EndFor
\EndProcedure
\For{each diagram D}
\State {Count(D,1)}
\EndFor
\end{algorithmic}
\end{algorithm}

\begin{figure}[H]
\centering
\includegraphics[width=0.6\textwidth]{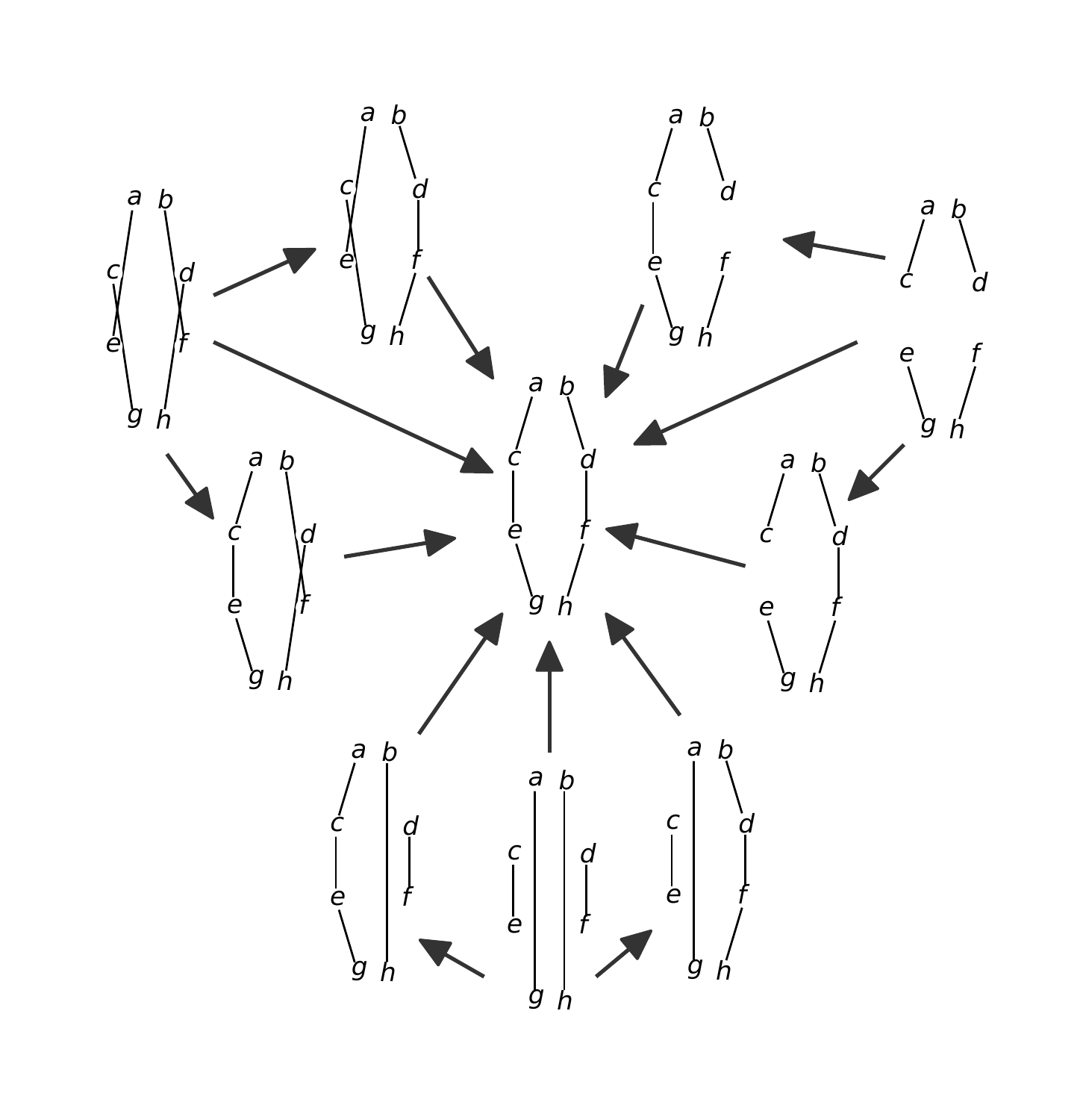}
\caption{Example of a subgraph of the graph of inclusion of the diagrams. Edges are directed toward the included diagrams. To assign a weight to a diagram, one has to start from this diagram in the graph and recursivelly subtract its neighbors as described in Algorithm 1.}
\label{}
\end{figure}

\subsubsection{Diagram types} 
Here we give one example for each diagram type

\begin{align}
    \diagramf{{ac,bd,eg,fh}}{a}=J_{ab}J_{cd}J_{ef}J_{gh}\delta_{ac}\delta_{bd}\delta_{eg}\delta_{fh}=J_{ab}J_{ab}J_{ef}J_{ef}=\Tr(J^2)^2
\end{align}
\begin{align}
    \diagramf{{ac,bd,eg,fh,df}}{a}=J_{ab}J_{cd}J_{ef}J_{gh}\delta_{ac}\delta_{bdfh}\delta_{fh}=J_{ab}J_{ab}J_{eb}J_{eb}=\sum_{b}\left((J^2)_{bb}\right)^2
\end{align}
\begin{align}
    \diagramf{{ac,bd,eg,fh,df,ce}}{a}=J_{ab}J_{cd}J_{ef}J_{gh}\delta_{aceg}\delta_{bdfh}=J_{ab}J_{ab}J_{ab}J_{ab}=\sum_{ab}(J_{ab})^4
\end{align}

\begin{align}
    \diagramf{{ac,eg,bh,df}}{a}=J_{ab}J_{cd}J_{ef}J_{gh}\delta_{ac}\delta_{bh}\delta_{df}\delta_{eg}=J_{ab}J_{ad}J_{ed}J_{eb}=\Tr J^4
\end{align}

\subsection{Result}
By procedurally enumerating all the terms and running the weighting algorithm, we get
\begin{align}
    \beta' = \frac{2^N}{16}\left(48\Tr J^4+12\Tr(J^2)^2-96 \sum_{a}\left((J^2)_{aa}\right)^2+32\sum_{ab}(J_{ab})^4\right)
\end{align}
so 
\begin{align}
    \beta &= \beta'+\beta''\\
    &=\frac{2^N}{16}\left(48\Tr J^4+12\Tr(J^2)^2-96 \sum_{a}\left((J^2)_{aa}\right)^2+32\sum_{ab}(J_{ab})^4\right)-\frac{2^N}{4}\left(\Tr J^2\right)^2\\
    &=2^N\left( 3 \Tr J^4+\frac{1}{2}\Tr(J^2)^2-6\sum_{a}\left((J^2)_{aa}\right)^2+2\sum_{ab}(J_{ab})^4\right)
\end{align}

The exact formula for $\lambda_2$ as defined in equation \eqref{lambda_2} if we only include the diagonal $\tau$'s is therefore
\begin{align}
    \lambda_2 &= \frac{\alpha}{2^N\beta}\\
    &=\frac{2\cdot \left(\Tr(J^4)-\sum_{e}\left[(J^2)_{ee}\right]^2\right)}{ 3 \Tr J^4+\frac{1}{2}\Tr(J^2)^2-6\sum_{a}\left((J^2)_{aa}\right)^2+2\sum_{ab}(J_{ab})^4}
\label{lamda_2J}
\end{align}
\begin{figure}
\centering
\includegraphics[width=0.6\textwidth]{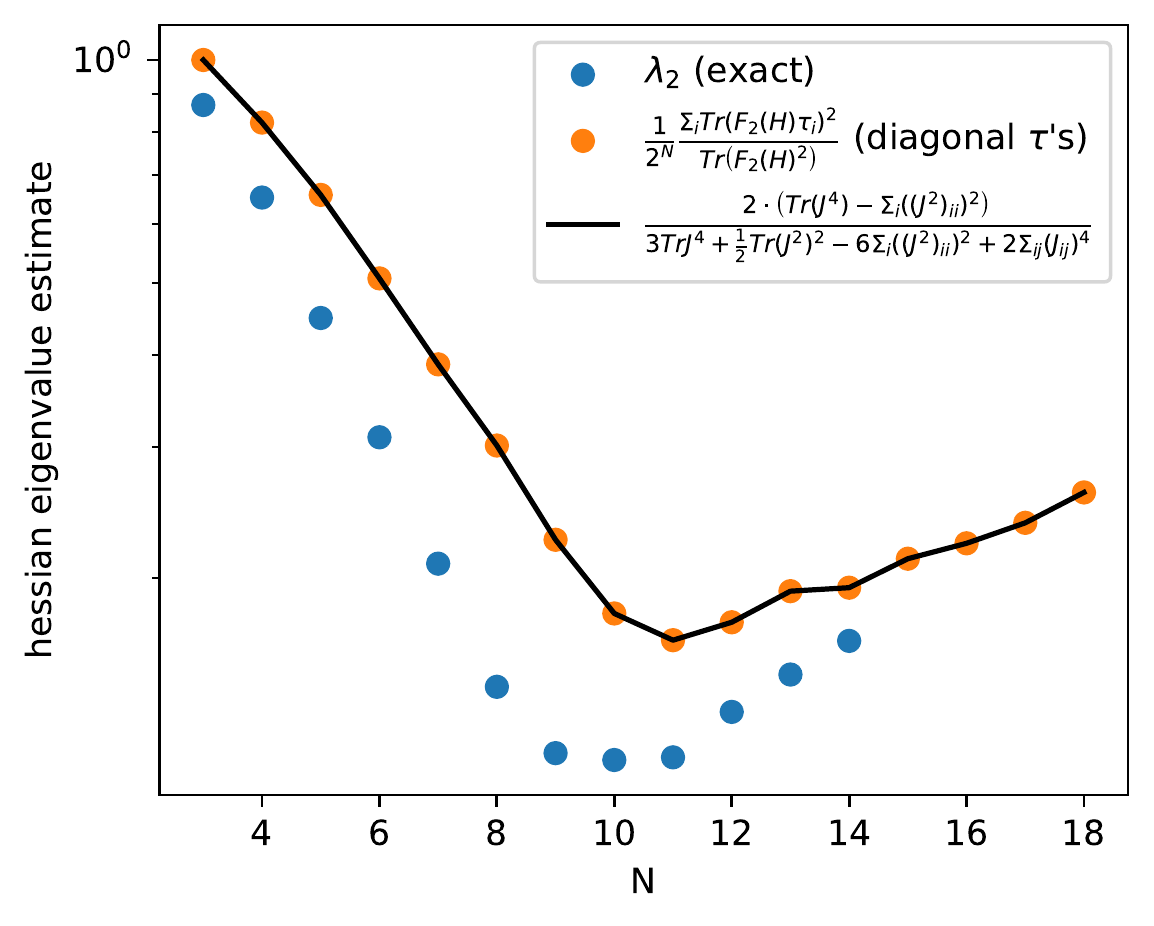}
\caption{We check our derivation by plotting equations \eqref{lambda_2} (orange dots) and \eqref{lamda_2J} (black line) versus $N$. Recall that this is an approximation of the exact $\lambda_2$ since these two formula only include diagonal $\tau$'s. The blue dots are the exact second eigenvalue computed by diagonalizing the hessian.}
\label{fig:lambda2}
\end{figure}

\section{SECTION 6 -- Low rank $J$ localization in the diagonal case}
\label{sup:lowrank}
\begin{figure}
\centering
\includegraphics[width=0.6\textwidth]{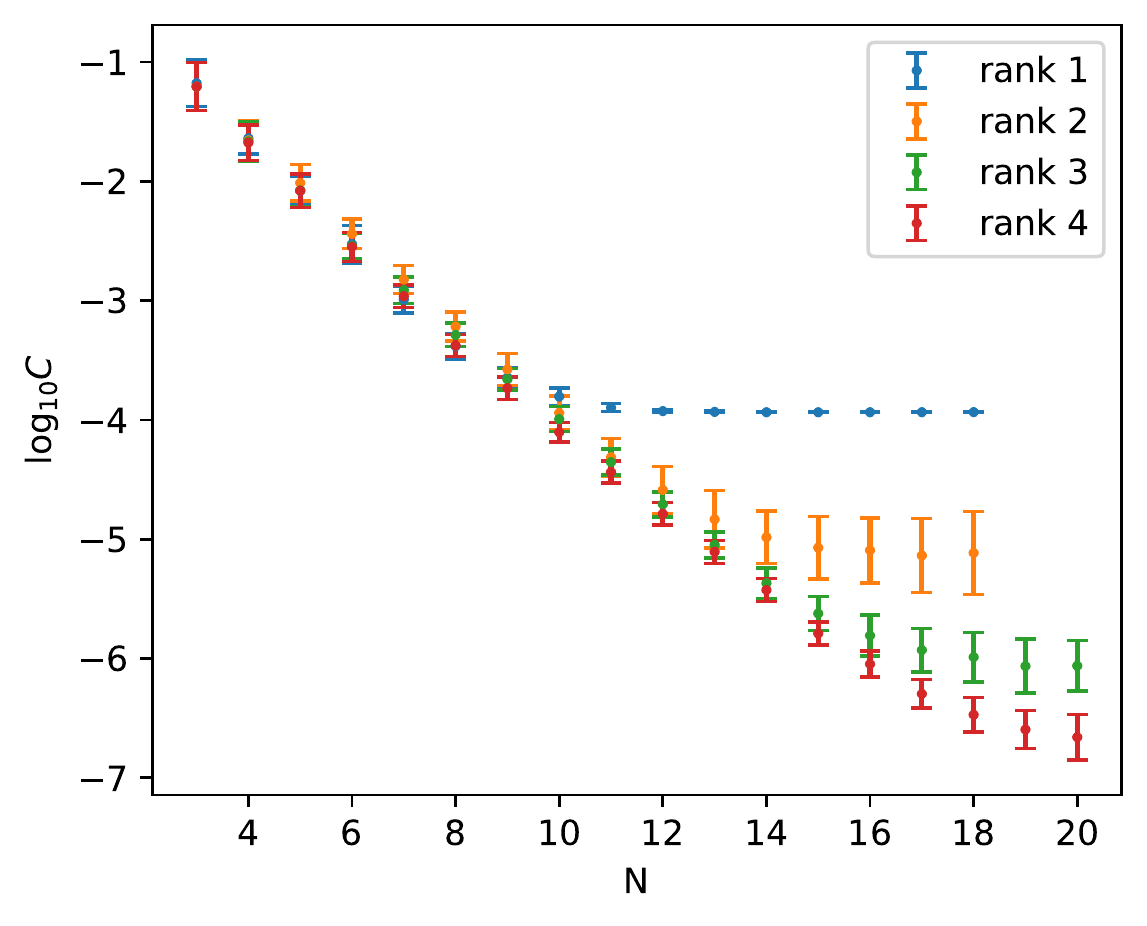}
\caption{Residual cost $C$ for two localization using only $ZZ$ strings versus $N$ for different truncations of the coupling matrix $J$. }
\label{fig:lowrank}
\end{figure}

In figure \ref{fig:lambda2}, we can see that $\lambda_2$ the quantity defined in equations \eqref{lamda_2J} and \eqref{lambda_2} increases after $N=11$. 
Note that in the large $N$ limit, equation \eqref{lamda_2J} becomes
\begin{align}
    \lambda_2 &\sim\frac{2 \Tr(J^4)}{ 3 \Tr J^4+\frac{1}{2}\Tr(J^2)^2}\\
    &\leq 4 \frac{ \Tr(J^4)}{\Tr(J^2)^2}
\end{align}
if $\lambda_2$ does not vanishes for large $N$, this suggests that $J$ low effective rank for large $N$ and that we could 2-localize GOE spectrum using only low rank $J$. We test this hypothesis in figure \ref{fig:lowrank}. Instead of minimising the cost function under the couplings $J$ (or weights $h_i$), we encode $J$ in a reduced number of eigenvectors $J=\sum_i v_i v^T_i$ and use the entries of the $v_i$ vectors as the parameters of the problem. For $1$ eigenvector, we find that this procedure is equivalent to 1-localization i.e the final cost after optimization is the same when trying to 1-localize or when trying to 2-localize with a rank-1 J.

\section{SECTION 7 -- Sparse localization}
\label{sup:sparse}
\begin{figure}[H]
\centering
\includegraphics[width=0.6\textwidth]{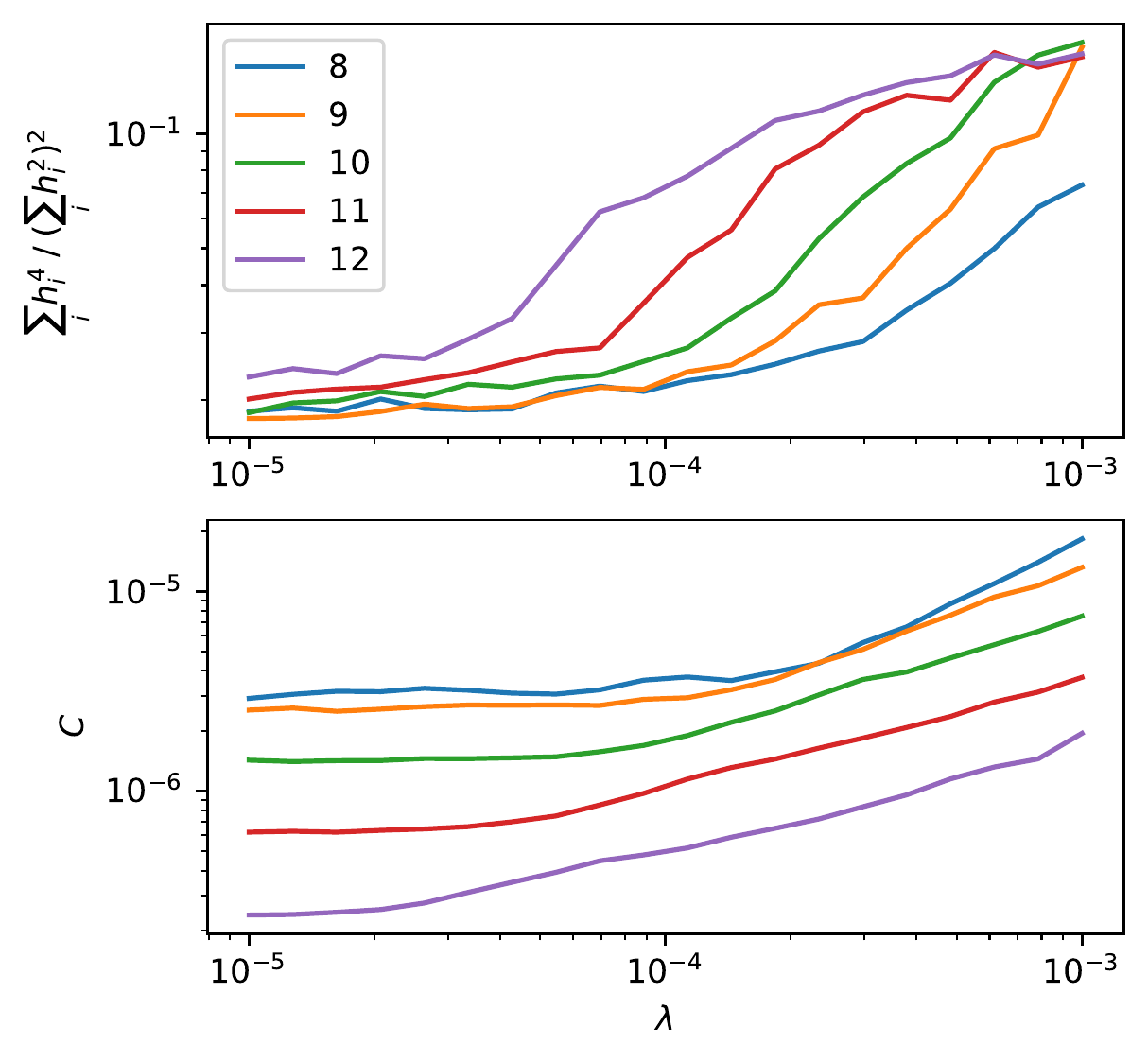}
\caption{Top: cost versus parameter $\lambda$. Note that this is the original cost, not including the $l1$ norm, but that the optimization has been performed using the $l1$ norm in the cost. Bottom: sparcity of $h$. A large value of $\sum_i h_i^4 \;/\; (\sum_i h_i^2)^2$ means that $h$ is sparse.}
\label{fig:sparse}
\end{figure}

There is no notion of geometric locality in the concept of 2-local Hamiltonian. One can force some geometry, e.g by forcing localization on a 1-d chain. Here we explore how sparse can the couplings $h_i$ be. Define a new cost function 
\begin{equation}
    C_\lambda(h) = \frac{1}{2^N}\sum_n(E_n-\Eps_n)^2+\lambda \sum |h_i|
\end{equation}
that enforce sparcity through $l1$ norm. In figure \ref{fig:sparse} we show how introducing this $l1$ norm in the cost impacts the sparcity of the result. We find that we can enforce significant sparcity while keeping the cost relavitelly low.

\end{document}